\documentclass[journal]{IEEEtran}
\usepackage{cite}
\usepackage{amsmath,amssymb}
\usepackage{graphicx}
\usepackage{booktabs}
\usepackage{siunitx}
\usepackage{xcolor}
\usepackage[hidelinks]{hyperref}

\providecommand{\BNOTRACKS}{---}

\newcommand{\BthreeCrsPd}{0.333}
\newcommand{\BthreeCrsPdSd}{0.021}
\newcommand{\BthreeCrsPdN}{5}

\newcommand{\BthreeCrsNObsMov}{59}

\newcommand{\BthreeCrsRangeCell}{2.44}

\newcommand{\BthreeCrsDopGateV}{0.2541}

\newcommand{\BthreeCrsCov}{78.3}
\newcommand{\BthreeCrsCovSd}{1.8}
\newcommand{\BthreeCrsCovN}{5}
\newcommand{\BthreeCrsCovNull}{26.7}

\newcommand{\BthreeCrsOspa}{15.69}

\newcommand{\BthreeCrsGospaLoc}{6.94}

\newcommand{\BthreeCrsIdf}{0.265}

\newcommand{\BthreeCrsIdsw}{21}

\newcommand{\BthreeCrsRmse}{7.68}

\newcommand{\BthreeCrsNis}{0.48}

\newcommand{\BthreeCrsCpis}{59}

\newcommand{\BthreeCrsCpiS}{0.64}

\newcommand{\BthreeCrsDecoyWorst}{0.0}

\newcommand{\BthreeCrsRawMean}{8.14}

\newcommand{\BthreeCrsMdp}{0.667}
\newcommand{\BthreeCrsMdpSd}{0.021}

\newcommand{\BthreeCrsFap}{0.515}

\newcommand{\BthreeCrsNReported}{41}

\newcommand{\BthreeCrsNUnassoc}{21}

\newcommand{\BthreeCrsSchedCpis}{59.4}

\newcommand{\BthreeCrsVisibleObs}{59.4}

\newcommand{\BthreeCrsRemovedPct}{0.00}

\newcommand{\BthreeCrsGain}{peak-hold, frozen}
\newcommand{\BthreeCrsTrim}{1.7}
\newcommand{\BthreeCrsCells}{5}
\newcommand{\BthreeCrsDecoyRatioUndef}{5}

\newcommand{\BthreeCrsDecoyMarginPp}{78.3}

\newcommand{\BthreeCrsMaxR}{85}

\newcommand{\BthreeCrsNullCov}{0.0}

\newcommand{\BthreeCrsNullOspa}{20.00}

\newcommand{\BthreeCrsNullGospaLoc}{0.00}

\newcommand{\BthreeCrsNullIdf}{0.000}

\newcommand{\BthreeCrsNullIdsw}{0}

\newcommand{\BthreeCrsNullNis}{\BNOTRACKS}

\newcommand{\BthreeCrsNullCpis}{61}

\newcommand{\BthreeCrsNullDecoyWorst}{0.0}

\newcommand{\BthreeCrsNullRawMean}{0.00}

\newcommand{\BthreeCrsBankedMdp}{0.666}

\newcommand{\BthreeCrsBankedFap}{0.571}

\newcommand{\BthreeSrsPd}{0.228}
\newcommand{\BthreeSrsPdSd}{0.024}
\newcommand{\BthreeSrsPdN}{5}

\newcommand{\BthreeSrsNObsMov}{56}

\newcommand{\BthreeSrsRangeCell}{2.44}

\newcommand{\BthreeSrsDopGateV}{0.2679}

\newcommand{\BthreeSrsCov}{73.0}

\newcommand{\BthreeSrsCovNull}{29.7}

\newcommand{\BthreeSrsOspa}{15.18}

\newcommand{\BthreeSrsGospaLoc}{7.24}

\newcommand{\BthreeSrsIdf}{0.383}

\newcommand{\BthreeSrsIdsw}{8}

\newcommand{\BthreeSrsRmse}{8.61}

\newcommand{\BthreeSrsNis}{0.49}

\newcommand{\BthreeSrsCpis}{56}

\newcommand{\BthreeSrsCpiS}{0.64}

\newcommand{\BthreeSrsDecoyWorst}{1.7}

\newcommand{\BthreeSrsRawMean}{5.78}

\newcommand{\BthreeSrsMdp}{0.772}
\newcommand{\BthreeSrsMdpSd}{0.024}

\newcommand{\BthreeSrsFap}{0.562}

\newcommand{\BthreeSrsNReported}{29}

\newcommand{\BthreeSrsNUnassoc}{16}

\newcommand{\BthreeSrsSchedCpis}{56.2}

\newcommand{\BthreeSrsVisibleObs}{56.0}

\newcommand{\BthreeSrsRemovedPct}{0.35}

\newcommand{\BthreeSrsRemovedPctWorst}{1.75}
\newcommand{\BthreeSrsGain}{pinned 82 dB}
\newcommand{\BthreeSrsTrim}{1.7}

\newcommand{\BthreeSrsDecoyMarginPp}{71.4}

\newcommand{\BthreeCrsAltPd}{0.351}
\newcommand{\BthreeCrsAltPdSd}{0.143}

\newcommand{\BthreeCrsAltCov}{4.0}
\newcommand{\BthreeCrsAltCovSd}{5.5}

\newcommand{\BthreeCrsAltCells}{5}

\newcommand{\BthreeCrsAltPass}{0}

\newcommand{\BthreeSrsNullCov}{0.0}

\newcommand{\BthreeSrsNullOspa}{20.00}

\newcommand{\BthreeSrsNullGospaLoc}{0.00}

\newcommand{\BthreeSrsNullIdf}{0.000}

\newcommand{\BthreeSrsNullIdsw}{0}

\newcommand{\BthreeSrsNullNis}{\BNOTRACKS}

\newcommand{\BthreeSrsNullCpis}{67}

\newcommand{\BthreeSrsNullDecoyWorst}{0.0}

\newcommand{\BthreeSrsNullRawMean}{0.00}

\newcommand{\BthreeHalfLMdp}{0.776}

\newcommand{\BthreeGateMidMdp}{0.667}

\newcommand{\BthreeGateMidFap}{0.692}

\newcommand{\BthreeGateMidMaxR}{200}

\newcommand{\BthreeGateWideFap}{0.780}

\newcommand{\BthreeGateWideMaxR}{315}

\newcommand{\BthreeRdmapCrsPresent}{+40.1}
\newcommand{\BthreeRdmapCrsAbsent}{+2.2}
\newcommand{\BthreeRdmapSrsPresent}{+48.4}
\newcommand{\BthreeRdmapSrsAbsent}{-4.7}
\newcommand{\BthreeRdmapCpi}{6}
\newcommand{\BthreeRdmapMarkR}{37.66}
\newcommand{\BthreeRdmapMarkV}{+2.32}
\newcommand{\BthreeRdmapPeakR}{39.0}
\newcommand{\BthreeRdmapRowLo}{19.5}
\newcommand{\BthreeRdmapRowHi}{56.1}

\newcommand{\BthreeSrsExecResolvePd}{0.02}
\newcommand{\BthreeSrsExecResolveCovPp}{3.6}

\newcommand{\BthreeSrsExecQuietN}{20}

\newcommand{\BthreeSrsExecQuietCpiMin}{52}
\newcommand{\BthreeSrsExecQuietCpiMax}{65}

\newcommand{\BthreeQuietSrsExecPdPooled}{0.0237}
\newcommand{\BthreeQuietSrsExecPdSeedPooled}{0.0205}

\newcommand{\BthreeQuietSrsExecCovPooled}{3.5614}
\newcommand{\BthreeQuietSrsExecCovSeedPooled}{3.0369}

\newcommand{\BthreeQuietSrsExecIdfPooled}{0.0984}
\newcommand{\BthreeQuietSrsExecIdfSeedPooled}{0.0867}

\newcommand{\BthreeArmDPdDtwo}{+0.105}
\newcommand{\BthreeArmDPdDtwoSe}{0.012}

\newcommand{\BthreeArmDPdDtwoPos}{5}
\newcommand{\BthreeArmDPdDtwoN}{5}
\newcommand{\BthreeArmDCovDtwo}{+5.2}

\newcommand{\BthreeArmDPdRep}{+0.073}
\newcommand{\BthreeArmDPdRepSe}{0.015}

\newcommand{\BthreeArmDPdRepPos}{5}
\newcommand{\BthreeArmDPdRepN}{5}
\newcommand{\BthreeArmDCovRep}{-2.9}

\newcommand{\MthirteenToneErrDb}{0.42}

\newcommand{\MthirteenReportedNzeroDbm}{-160.0}

\newcommand{\MthirteenFloorGapDb}{-67.6}

\newcommand{\MthirteenBwMHz}{18.0}
\newcommand{\MthirteenNfDb}{9.0}

\newcommand{\MthirteenDigitalNoiseLsb}{1.43}
\newcommand{\MthirteenNzeroOrigin}{lte\_ue\_measurements.c:264, gain- and FFT-referred, printed as an integer}

\newcommand{\DiffLoCovVone}{13.7}
\newcommand{\DiffHiCovVone}{72.9}
\newcommand{\DiffLoCovVtwo}{0.0}
\newcommand{\DiffHiCovVtwo}{4.1}

\newcommand{\DiffCrsShiftPp}{-16.7}

\newcommand{\DiffSrsShiftPp}{-19.4}

\newcommand{\SimTwoTapCount}{2}

\newcommand{\SimTwoTapRmsM}{0.11}

\newcommand{\LimGateM}{20}
\newcommand{\LimGateWideM}{30}
\newcommand{\LimModeHi}{84}
\newcommand{\LimModeLo}{68}

\newcommand{\LimCeil}{87.8}
\newcommand{\LimConvLo}{87.5}
\newcommand{\LimConvHi}{87.8}

\newcommand{\LimElemBwAzDeg}{65}
\newcommand{\LimElemBwElDeg}{65}
\newcommand{\LimPanelBwLoDeg}{6}
\newcommand{\LimPanelBwHiDeg}{7}

\newcommand{\LimAdcBits}{16}
\newcommand{\LimThermalBwMHz}{18}

\newcommand{\LimNoiseRmsLsb}{0.12}

\newcommand{\LimDirectHeadroomDb}{3.0}

\newcommand{\LimFieldDirectOverNoiseDb}{69}
\newcommand{\PanelRatioFifteenLoDb}{-1.93}
\newcommand{\PanelRatioFifteenHiDb}{+1.10}
\newcommand{\PanelRatioSixtyLoDb}{-21.66}
\newcommand{\PanelRatioSixtyHiDb}{-6.32}
\newcommand{\LimAerialZ}{60}

\newcommand{\LimFacingLiftPp}{+6.29}
\newcommand{\LimFacingLiftSePp}{3.40}

\newcommand{\TaPfaOn}{6.6e-6}
\newcommand{\TaPfaOff}{2.0e-6}
\newcommand{\TaGainLoPct}{62}
\newcommand{\TaGainHiPct}{76}

\newcommand{\SpeedThreeMdp}{0.780}

\newcommand{\SpeedSevenMdp}{0.556}

\newcommand{\LtVehN}{5}
\newcommand{\LtVehIDF}{0.900}
\newcommand{\LtVehIDFSd}{0.039}
\newcommand{\LtVehGospa}{6.28}
\newcommand{\LtVehGospaSd}{0.39}
\newcommand{\LtVehPd}{0.551}
\newcommand{\LtVehPdSd}{0.013}

\newcommand{\LtVehIdsw}{1.20}
\newcommand{\LtVehIdswZero}{4}

\newcommand{\LtVehNullCov}{0.00}
\newcommand{\LtUavAN}{5}
\newcommand{\LtUavAIDF}{0.761}
\newcommand{\LtUavAIDFSd}{0.050}
\newcommand{\LtUavAGospa}{8.84}
\newcommand{\LtUavAGospaSd}{1.08}
\newcommand{\LtUavAPd}{0.632}
\newcommand{\LtUavAPdSd}{0.006}

\newcommand{\LtUavAIdsw}{1.80}

\newcommand{\LtUavBN}{5}
\newcommand{\LtUavBIDF}{0.796}
\newcommand{\LtUavBIDFSd}{0.081}
\newcommand{\LtUavBGospa}{6.50}
\newcommand{\LtUavBGospaSd}{0.31}
\newcommand{\LtUavBPd}{0.563}
\newcommand{\LtUavBPdSd}{0.014}

\newcommand{\LtUavBIdsw}{1.40}

\newcommand{\LtLapM}{283}
\newcommand{\LtLapSpeed}{4.60}

\newcommand{\LtVehZ}{1.6}
\newcommand{\LtUavAZ}{15}
\newcommand{\LtUavBZ}{30}

\newcommand{\LtZVehN}{15}
\newcommand{\LtZVehIDF}{0.926}

\newcommand{\LtZVehPd}{0.315}
\newcommand{\LtZVehPdSd}{0.004}

\newcommand{\LtZUavAN}{21}
\newcommand{\LtZUavAIDF}{0.690}

\newcommand{\LtZUavAPd}{0.394}
\newcommand{\LtZUavAPdSd}{0.007}

\newcommand{\LtZUavBN}{20}
\newcommand{\LtZUavBIDF}{0.738}

\newcommand{\LtZUavBPd}{0.322}
\newcommand{\LtZUavBPdSd}{0.007}

\newcommand{\HstSlowOnN}{15}

\newcommand{\HstSlowOnCpis}{27}
\newcommand{\HstSlowOnNdet}{8.1}

\newcommand{\HstEnbX}{531.0}
\newcommand{\HstEnbY}{-747.1}
\newcommand{\HstEnbZ}{29.0}
\newcommand{\HstUeX}{287.3}
\newcommand{\HstUeY}{-547.7}
\newcommand{\HstUeZ}{1.5}
\newcommand{\HstBaseline}{316}
\newcommand{\HstLapLen}{105}
\newcommand{\HstSpeed}{5.47}
\newcommand{\HstSpeedKmh}{20}
\newcommand{\HstLapS}{19.2}
\newcommand{\HstCpiDt}{0.64}
\newcommand{\HstFcMHz}{1845}
\newcommand{\HstBwMHz}{20}
\newcommand{\HstPoolN}{15}
\newcommand{\HstPoolCov}{48.1}
\newcommand{\HstPoolCovSd}{14.2}

\newcommand{\HstPoolPass}{10}
\newcommand{\HstPoolCovPass}{55.4}

\newcommand{\HstNullN}{5}

\newcommand{\HstSitAN}{15}
\newcommand{\HstSitACov}{42.3}
\newcommand{\HstSitACovSd}{19.1}

\newcommand{\HstSitAPass}{6}

\newcommand{\HstSitASub}{3.7}

\newcommand{\HstSitABrN}{15}
\newcommand{\HstSitABrCov}{32.9}
\newcommand{\HstSitABrCovSd}{26.2}

\newcommand{\HstSitABrPass}{9}

\newcommand{\HstSitBN}{15}
\newcommand{\HstSitBCov}{48.1}
\newcommand{\HstSitBCovSd}{14.2}

\newcommand{\HstSitBPass}{10}

\newcommand{\HstSitBSub}{5.9}

\newcommand{\HstSitCN}{15}
\newcommand{\HstSitCCov}{42.7}
\newcommand{\HstSitCCovSd}{11.3}

\newcommand{\HstSitCPass}{6}

\newcommand{\HstSitCSub}{10.5}

\newcommand{\HstSitDN}{15}
\newcommand{\HstSitDCov}{46.1}
\newcommand{\HstSitDCovSd}{15.1}

\newcommand{\HstSitDPass}{13}

\newcommand{\HstSitDSub}{14.6}

\newcommand{\HstSitEN}{15}
\newcommand{\HstSitECov}{52.4}
\newcommand{\HstSitECovSd}{8.6}

\newcommand{\HstSitEPass}{14}

\newcommand{\HstSitESub}{21.3}

\newcommand{\HstSitFN}{15}
\newcommand{\HstSitFCov}{61.7}
\newcommand{\HstSitFCovSd}{19.4}

\newcommand{\HstSitFPass}{14}

\newcommand{\HstSitFSub}{27.8}

\newcommand{\HstSitGN}{15}
\newcommand{\HstSitGCov}{60.9}
\newcommand{\HstSitGCovSd}{11.6}

\newcommand{\HstSitGPass}{14}

\newcommand{\HstSitGSub}{40.3}

\newcommand{\HstSitHN}{15}
\newcommand{\HstSitHCov}{29.9}
\newcommand{\HstSitHCovSd}{11.7}

\newcommand{\HstSitHPass}{9}

\newcommand{\HstSitHSub}{88.0}

\newcommand{\HstSitIN}{15}
\newcommand{\HstSitICov}{32.3}
\newcommand{\HstSitICovSd}{14.1}

\newcommand{\HstSitIPass}{5}

\newcommand{\HstSitISub}{121.8}

\newcommand{\HstSitIBrN}{15}
\newcommand{\HstSitIBrCov}{22.7}
\newcommand{\HstSitIBrCovSd}{25.0}

\newcommand{\HstSitIBrPass}{5}

\newcommand{\MpCoverA}{34.5}
\newcommand{\MpVdopGroundA}{16.6}
\newcommand{\MpVdopAirA}{1.65}
\newcommand{\MpCoverD}{69.5}
\newcommand{\MpVdopGroundD}{5.95}
\newcommand{\MpVdopAirD}{1.17}
\newcommand{\MpCoverC}{100.0}
\newcommand{\MpVdopGroundC}{5.66}
\newcommand{\MpVdopAirC}{0.95}

\newcommand{\MpVdopAirAflat}{1.67}
\newcommand{\MpCoChannelSectors}{14}
\newcommand{\MpSites}{5}
\newcommand{\MpAirZ}{100}
\newcommand{\MpResyncN}{475}
\newcommand{\MpResyncIds}{4}

\newcommand{\HstSiteN}{7}
\newcommand{\HstSiteRecN}{71}
\newcommand{\HstNbrN}{3}
\newcommand{\HstTxSid}{102006}
\newcommand{\HstTxLicFc}{1812.5}
\newcommand{\HstTxLicBw}{15}

\newcommand{\HaTxSid}{303268}
\newcommand{\HaTxX}{-130.9}
\newcommand{\HaTxY}{-828.0}
\newcommand{\HaTxZ}{15.3}
\newcommand{\HaUeX}{-77.7}
\newcommand{\HaUeY}{-578.0}
\newcommand{\HaUeZ}{8.0}
\newcommand{\HaBaseline}{255.7}
\newcommand{\HaLapM}{137}
\newcommand{\HaSpeedMax}{3.5}
\newcommand{\HaFcMHz}{1842.5}
\newcommand{\HaStreetBearing}{7.7}
\newcommand{\HaTxOffCentreM}{17.6}
\newcommand{\HaSubDeg}{0.7}
\newcommand{\HaRefSubDeg}{14.6}
\newcommand{\HaRhoDotOverV}{1.998}

\newcommand{\HaN}{5}
\newcommand{\HaVehZ}{1.6}
\newcommand{\HaVehCov}{85.36}
\newcommand{\HaVehCovSd}{0.98}
\newcommand{\HaVehPerm}{41.36}
\newcommand{\HaVehPermSd}{6.90}
\newcommand{\HaVehMargin}{44.00}
\newcommand{\HaVehMarginSd}{6.83}
\newcommand{\HaVehMarginLo}{+35.5}
\newcommand{\HaVehMarginHi}{+49.6}

\newcommand{\HaVehPd}{0.381}
\newcommand{\HaVehPdSd}{0.021}
\newcommand{\HaVehIDF}{0.175}
\newcommand{\HaVehIDFSd}{0.055}
\newcommand{\HaVehIdswLo}{11}
\newcommand{\HaVehIdswHi}{19}
\newcommand{\HaVehRmse}{7.50}
\newcommand{\HaVehRmseSd}{1.12}
\newcommand{\HaVehNullCov}{0.0}
\newcommand{\HaUavZ}{15}
\newcommand{\HaUavCov}{84.98}
\newcommand{\HaUavCovSd}{1.57}
\newcommand{\HaUavPerm}{41.54}
\newcommand{\HaUavPermSd}{1.89}
\newcommand{\HaUavMargin}{43.44}
\newcommand{\HaUavMarginSd}{2.09}
\newcommand{\HaUavMarginLo}{+41.6}
\newcommand{\HaUavMarginHi}{+46.5}

\newcommand{\HaUavPd}{0.396}
\newcommand{\HaUavPdSd}{0.006}
\newcommand{\HaUavIDF}{0.159}
\newcommand{\HaUavIDFSd}{0.027}
\newcommand{\HaUavIdswLo}{13}
\newcommand{\HaUavIdswHi}{18}
\newcommand{\HaUavRmse}{5.75}
\newcommand{\HaUavRmseSd}{0.50}
\newcommand{\HaUavNullCov}{0.0}
\newcommand{\HaRefCov}{85.02}
\newcommand{\HaRefCovSd}{3.84}
\newcommand{\HaRefPerm}{13.32}
\newcommand{\HaRefPermSd}{0.70}
\newcommand{\HaRefMargin}{71.70}
\newcommand{\HaRefMarginSd}{3.41}
\newcommand{\HaRefMarginLo}{+66.7}
\newcommand{\HaRefMarginHi}{+75.0}

\newcommand{\HaRefPd}{0.521}
\newcommand{\HaRefPdSd}{0.048}
\newcommand{\HaRefIDF}{0.248}
\newcommand{\HaRefIDFSd}{0.013}
\newcommand{\HaRefIdswLo}{3}
\newcommand{\HaRefIdswHi}{14}
\newcommand{\HaRefRmse}{6.35}
\newcommand{\HaRefRmseSd}{0.92}
\newcommand{\HaRefNullCov}{2.0}
\newcommand{\HaRefCovTaZero}{54.84}
\newcommand{\HaRefCovTaZeroSd}{5.45}

\newcommand{\HaCarrierDeltaPp}{-2.38}
\newcommand{\HaCarrierDeltaT}{-1.32}
\newcommand{\HcSid}{102006}
\newcommand{\HcFcMHz}{1812.5}
\newcommand{\HcDistM}{632.1}
\newcommand{\HcOffBoreDeg}{44.5}

\newcommand{\HcDirectRelDb}{-10.73}
\newcommand{\HcRatioRelDb}{-2.52}
\newcommand{\HcLosSnap}{16}
\newcommand{\HcSnapN}{62}
\newcommand{\HcLosPct}{26}
\newcommand{\HcDet}{1}
\newcommand{\HcObs}{16}
\newcommand{\HcPd}{0.062}
\newcommand{\HcCov}{0.0}
\newcommand{\HcPerm}{0.0}

\newcommand{\HcWorstDecoy}{0.0}

\newcommand{\HcNdet}{307}

\newcommand{\HcNdetPrimary}{164}
\newcommand{\HcNdetPrimarySd}{12}
\newcommand{\HcNdetPrimaryN}{5}

\newcommand{\HoOffsetM}{15}
\newcommand{\HoSubDeg}{44.5}
\newcommand{\HoEchoDb}{-2.57}
\newcommand{\HoN}{5}
\newcommand{\HoCov}{77.78}
\newcommand{\HoCovSd}{13.67}

\newcommand{\HoPd}{0.365}
\newcommand{\HoPdSd}{0.065}
\newcommand{\HoIDF}{0.211}
\newcommand{\HoIDFSd}{0.076}

\newcommand{\HoIdsw}{9.6}
\newcommand{\HoIdswSd}{2.7}
\newcommand{\HoMargin}{39.64}
\newcommand{\HoMarginSd}{10.77}
\newcommand{\HoRefIdsw}{15.2}
\newcommand{\HoRefIdswSd}{4.0}

\newcommand{\HoDRmse}{+1.41}
\newcommand{\HoDRmsePos}{5}

\newcommand{\HoOutlierCov}{53.8}

\newcommand{\HcxN}{8}
\newcommand{\HcxGood}{8}

\newcommand{\HcxSpacingM}{135}
\newcommand{\HcxStretchM}{155}
\newcommand{\HcxOverlapM}{20}
\newcommand{\HcxSpanM}{1140}
\newcommand{\HcxLapLenM}{195}
\newcommand{\HcxYfirst}{-578}
\newcommand{\HcxYlast}{367}
\newcommand{\HcxDirectLoDb}{-10.57}
\newcommand{\HcxRatioLoDb}{7.3}
\newcommand{\HcxRatioHiDb}{7.8}
\newcommand{\HcxVehCovLo}{44.2}
\newcommand{\HcxVehCovHi}{83.2}
\newcommand{\HcxVehCovW}{69.4}

\newcommand{\HcxUavCovW}{68.3}
\newcommand{\HcxNullLo}{0.0}
\newcommand{\HcxNullHi}{20.9}
\newcommand{\HcxNullAboveFive}{5}

\newcommand{\HcxSeedsLo}{5}

\newcommand{\HcxVehDem}{5}
\newcommand{\HcxUavDem}{4}
\newcommand{\HcxVehCells}{40}
\newcommand{\HcxUavCells}{40}
\newcommand{\HcxVehK}{37}
\newcommand{\HcxUavK}{32}
\newcommand{\HcxVehSdAllPassLo}{1.6}
\newcommand{\HcxVehSdAllPassHi}{7.7}
\newcommand{\HcxVehSdLo}{1.6}
\newcommand{\HcxVehSdHi}{30.4}

\newcommand{\HcxBimodalRx}{2,3,7,8}
\newcommand{\HcxVehBimodalRx}{1,2,7}
\newcommand{\HcxVehCovPassLo}{45.9}
\newcommand{\HcxVehCovPassHi}{83.3}
\newcommand{\HcxUavCovPassLo}{40.4}
\newcommand{\HcxUavCovPassHi}{84.7}

\newcommand{\HcxVehPdLo}{0.17}
\newcommand{\HcxVehPdHi}{0.33}
\newcommand{\HcxUavPdLo}{0.16}
\newcommand{\HcxUavPdHi}{0.34}

\newcommand{\LbEirpDbm}{65.83}
\newcommand{\LbRcsDbsm}{10}
\newcommand{\LbNfDb}{9}
\newcommand{\LbThermalDbm}{-92.42}
\newcommand{\LbCohGainDb}{41.07}
\newcommand{\LbReN}{200}
\newcommand{\LbOccN}{64}
\newcommand{\LbRxEightBaselineM}{1209}
\newcommand{\LbRxEightDirDbm}{-33.6}
\newcommand{\LbRxEightEchoEndDbm}{-80.7}
\newcommand{\LbRxEightSnrEndDb}{52.8}
\newcommand{\LbRxEightMarginDb}{51.1}
\newcommand{\LbRxEightDpiOverEchoDb}{47}

\newcommand{\GfGateM}{400}
\newcommand{\GfRx}{3}
\newcommand{\GfCovThree}{83.2}
\newcommand{\GfCovThreeSd}{1.6}
\newcommand{\GfKThree}{5}
\newcommand{\GfCovFour}{67.6}
\newcommand{\GfCovFourSd}{25.3}
\newcommand{\GfKFour}{4}
\newcommand{\GfPdThree}{0.328}
\newcommand{\GfPdThreeSd}{0.023}
\newcommand{\GfPdFour}{0.401}
\newcommand{\GfPdFourSd}{0.121}
\newcommand{\GfRepThree}{292}
\newcommand{\GfRepFour}{619}
\newcommand{\GfJointBandM}{21}
\newcommand{\GfMissM}{4}
\newcommand{\GfCoinc}{21}
\newcommand{\GfOneRxM}{1.46}
\newcommand{\GfOneRxSdM}{1.13}
\newcommand{\GfTwoRxM}{1.54}
\newcommand{\GfTwoRxSdM}{1.06}
\newcommand{\GfPermLoM}{2.1}
\newcommand{\GfPermHiM}{9.7}
\newcommand{\GfRangeCellProjM}{1.03}

\newcommand{\BthreeCrsGridFsMHz}{30.72}
\newcommand{\BthreeCrsGridSampleM}{9.76}

\newcommand{\BthreeCrsGridRfft}{8192}
\newcommand{\BthreeCrsGridPad}{4}
\newcommand{\BthreeCrsGridDeltaRM}{2.44}
\newcommand{\BthreeCrsGridNDop}{64}
\newcommand{\BthreeCrsGridDeltaVMs}{0.2539}

\newcommand{\BthreeCrsGridRangeTaper}{Hann}
\newcommand{\BthreeCrsGridDopplerTaper}{Hamming}

\newcommand{\BthreeCrsPilotApWindowedBwMHz}{17.94}
\newcommand{\BthreeCrsPilotApCoverBM}{16.71}

\newcommand{\BthreeCrsPfTwoCellMean}{0.5149}
\newcommand{\BthreeCrsPfTwoReport}{0.5194}

\newcommand{\BthreeCrsPfTwoNInstances}{206}

\newcommand{\BthreeCrsPfTwoMaxObjects}{1}

\newcommand{\BthreeCrsPfOneCpis}{63}
\newcommand{\BthreeCrsPfOneObjects}{0}
\newcommand{\BthreeCrsPfOne}{0.0000}
\newcommand{\BthreeCrsPfOneRawCpis}{2}

\newcommand{\BthreeSrsPfOneCpis}{68}
\newcommand{\BthreeSrsPfOneObjects}{0}
\newcommand{\BthreeSrsPfOne}{0.0000}
\newcommand{\BthreeSrsPfOneRawCpis}{1}

\newcommand{\BthreeCrsContRunsPerCell}{1.4}
\newcommand{\BthreeCrsContLongestFrames}{295}
\newcommand{\BthreeCrsContLongestS}{206.5}
\newcommand{\BthreeCrsContLongestM}{90.1}

\newcommand{\BthreeCrsContIdLongestM}{74.1}

\newcommand{\BthreeSrsContRunsPerCell}{2.0}

\newcommand{\BthreeSrsContLongestM}{83.6}

\begin{document}

\title{Vehicle and UAV Detection and Tracking with LTE Downlink Reference
Signals: A Ray-Traced Evaluation on a Mapped Suburban Site with Licensed
Base-Station Records}
\author{Arun~K.~Gurung,~\IEEEmembership{Senior Member,~IEEE,}
and~Satha~K.~Sathananthan%
\thanks{A. K. Gurung is with Krasara Pty Ltd, Australia (e-mail:
arun.gurung@krasara.com.au).}%
\thanks{S. K. Sathananthan is with Nexvis Pty Ltd, Australia (e-mail:
satha@nexvis.com.au).}%
\thanks{Manuscript revised 2026-11-01.}}
\markboth{Preprint}{Gurung and Sathananthan: Vehicle and UAV Detection and Tracking with LTE Downlink Reference Signals on a Mapped Site}
\maketitle

\begin{abstract}
Passive sensing on the cellular downlink places the receiver at the user
equipment and changes nothing in the network. We report simulation results
for an LTE band-3 cell-specific reference signal arm on a ray-traced channel. On a
synthetic street canyon the arm is compared with an uplink sounding reference
signal arm on the same band, target and scorer: missed-detection probabilities
of $\BthreeCrsMdp \pm \BthreeCrsMdpSd$ and
$\BthreeSrsMdp \pm \BthreeSrsMdpSd$. On a surveyed site
built from OpenStreetMap footprints and the licensed site record, nine
receiver positions along the street track the vehicle at coverage between
\SI{\HstSitHCov}{\percent} and \SI{\HstSitFCov}{\percent}; a
\SI{\LtLapM}{\metre} lap separates target height from trajectory at
\SI{\LtVehZ}{} to \SI{\LtUavBZ}{\metre}; a \SI{\HaLapM}{\metre} lap on the
built-up street tracks a vehicle and a \SI{\HaUavZ}{\metre} airborne target on
every cell, and \HcxN{} receivers at \SI{\HcxSpacingM}{\metre} spacing carry
one illuminator's coverage over \SI{\HcxSpanM}{\metre} of the street. Potential
future work includes extending the analysis to multiple illuminators and varying
receiver configurations, and confirmation through over-the-air measurements.
\end{abstract}


\section{Introduction}
A passive cellular sensing receiver observes a target through the echo of a
signal transmitted for communication. In the downlink configuration studied here it sits
at the user-equipment end of the link, which requires no change
at the base station. On the uplink it sits at the base station, which is
where the processing, the calibration and the timing reference already are.

Results for the two are usually reported separately, on different bands, scenes
and scorers. The receiving node, the reference signal, the bandwidth and the
association gate then vary together, and a difference between two published
numbers cannot be attributed to any one of them. This paper reports the two
arms on one band, one scene, one trajectory and one scorer, so that the
remaining differences are enumerated.

\textbf{Contributions.} (i) A matched-band downlink and uplink passive sensing
comparison on LTE band~3, scored by one scorer at one association gate, with
every unequalised item enumerated in Section~\ref{sec:arms}. (ii) Detection
reported in the 3GPP sensing quantities alongside the in-house tracking
quantities, with the mapping between them stated in
Section~\ref{sec:metrics}. (iii) A null discipline in which every arm is
reported against a target-absent replay of its own scene and against
displaced-truth nulls, so that a coverage figure is read against the coverage the
same pipeline reports for a displaced truth. (iv) The downlink arm carried to a
surveyed site built from public building footprints and the national licence
record (Section~\ref{sec:highst}): receiver siting, trajectory length, target
height from a ground vehicle to a \SI{\LtUavBZ}{\metre} airborne target, and a
lap on the built-up street itself, each with the same nulls. (v) One
illuminator's coverage extended along \SI{\HcxSpanM}{\metre} of the street
by a chain of receivers run one at a time and assembled by time ordering
(Section~\ref{sec:chain}). (vi) The cells licensed at the site catalogued as
further illuminators for one receiver, with their geometric dilution and one
pair measured (Section~\ref{sec:multipci}).

Table~\ref{tab:acronyms} lists the acronyms used; the metrics are defined in
Section~\ref{sec:metrics}.

\begin{table}[t]
\caption{Acronyms.}
\label{tab:acronyms}
\centering
\footnotesize
\setlength{\tabcolsep}{3pt}
\begin{tabular}{@{}ll@{}}
\toprule
ACMA & Australian Communications and Media Authority \\
CFAR & constant false-alarm rate (OS: ordered-statistic) \\
CPI & coherent processing interval \\
CRS & cell-specific reference signal (LTE downlink) \\
eNB & evolved Node B, the LTE base station \\
EIRP & equivalent isotropically radiated power \\
FAP\textsubscript{2} & type-2 false-alarm probability \\
FDD & frequency-division duplex \\
GOSPA & generalized optimal sub-pattern assignment metric \\
IDF1, IDSW & identification F1 score; identity switches \\
ISAC & integrated sensing and communication \\
LOS & line of sight \\
LTE & Long-Term Evolution (4G cellular) \\
MDP & missed-detection probability \\
NIS & normalised innovation squared \\
O-RAN & Open Radio Access Network \\
OSM & OpenStreetMap \\
PCI & physical cell identity \\
PRB & physical resource block \\
$P_\mathrm{D}$ & probability of detection \\
RCS & radar cross-section \\
RMSE & root-mean-square error \\
SCS & subcarrier spacing \\
SRS & sounding reference signal (LTE uplink) \\
UAV & unmanned aerial vehicle \\
UE & user equipment \\
VDOP & vertical dilution of precision \\
\bottomrule
\end{tabular}
\end{table}

\textbf{Scope.} The uplink arm is a baseline for the downlink arm. The channel is simulated. Each scene carries one target. The comparison of
Sections~\ref{sec:arms}--\ref{sec:variance} is between two receiving geometries
at one operating point.

\section{Related work and 3GPP context}
Integrated sensing and communication is a 3GPP study item, and the service
requirements and use cases for it are set out in \cite{3gpp-tr22837}. The two
reference signals used here are standardised: the LTE cell-specific reference
signal in \cite{3gpp-ts36211}, which is transmitted continuously on the downlink
and requires no scheduling, and the sounding reference signal in
\cite{3gpp-ts36211}, which is transmitted by the user equipment on the uplink.
The channel is a ray-traced trace of the scene geometry, computed with an
open-source differentiable ray tracer~\cite{sionna,sionna-rt}; the array element
patterns follow \cite{3gpp-tr38901}.

Passive sensing on broadcast downlink signals and on uplink sounding signals are
both established lines.
Passive radar on the LTE downlink is reviewed in \cite{rai-lte-passive-review-2021};
the \SI{10}{\milli\second} frame periodicity of the cell-specific reference signal
and the inter-frame ambiguity it produces, together with its suppression by
per-symbol normalisation of the reference, are analysed in
\cite{wang-lte-interframe-2018}. The closest experimental comparisons are
LTE passive bistatic detection of ground-moving targets over the air
\cite{raja-abdullah-lte-pbr-2016}, a multistatic LTE system for UAV detection
that proposes a dedicated localisation reference signal rather than the
unmodified cell-specific one used here \cite{geng-lte-multistatic-uav-2020},
and two recent experimental studies of passive UAV tracking on cellular
downlink signals, one with a digital array measuring range, Doppler and angle
of arrival \cite{sun-passive-uav-tracking-2024} and one reconstructing a drone
trajectory from Doppler at three passive receivers
\cite{ji-doppler-multistatic-drone-2025}, with a follow-on bistatic study
\cite{sun-bistatic-uav-2026}. Table~\ref{tab:relatedwork} places this paper
beside them on the dimensions a reader needs to compare results: every
comparison work measures a real channel over the air, and this paper does not;
this paper reports controls with a known answer and a scoring convention held
across every table, which the comparison works, as published, do not report.

\begin{table*}[t]
\caption{This paper beside the closest LTE passive-sensing experiments.
Entries are taken from the cited abstracts and texts; ``n.r.'' is not reported
there. The comparison works measure over the air; this paper is simulation
only.}
\label{tab:relatedwork}
\centering
\scriptsize
\setlength{\tabcolsep}{3pt}
\begin{tabular}{@{}p{1.8cm}p{1.9cm}p{1.8cm}p{1.6cm}p{1.6cm}p{1.8cm}p{1.5cm}p{2.2cm}p{1.9cm}@{}}
\toprule
work & waveform used & channel & receive array & synchronisation & targets & clutter & tracking metrics & published controls \\
\midrule
\cite{raja-abdullah-lte-pbr-2016} & LTE downlink, full signal & measured, over the air & two channels (reference, surveillance) & reference channel & ground vehicles, six scenarios & real, direct-path cancellation & detection only & n.r. \\
\cite{geng-lte-multistatic-uav-2020} & LTE with a proposed localisation reference signal & measured and simulated & multiple receivers & network & UAV & real & localisation error & n.r. \\
\cite{sun-passive-uav-tracking-2024} & LTE downlink & measured, over the air & digital array & direct path & UAV & real & range, Doppler, AoA, tracking error (metre level) & n.r. \\
\cite{ji-doppler-multistatic-drone-2025} & LTE downlink & measured, over the air & three single-channel receivers & direct path & drone & real & Doppler, trajectory error & n.r. \\
this paper & LTE CRS only (and SRS on the uplink arm) & ray-traced, two taps, direct path normalised & $1\times 2$ & self-referenced to the cell's direct path & one surrogate sphere; vehicle and \SI{15}{\metre} airborne & rendered geometry only & $P_D$, MDP, FAP$_2$, coverage, IDF1, IDSW, OSPA/GOSPA, RMSE, NIS & permutation null, decoy null, target-absent null, execution repeat, one convention \\
\bottomrule
\end{tabular}
\end{table*}
The uplink arm descends from an end-to-end O-RAN testbed in which the 5G NR
sounding reference signal is used unmodified as the sensing waveform, detections
are tracked by an extended Kalman filter deployed as an xApp, and the elevation
ambiguity of a single bistatic pair is resolved by a vertical receive array or a
second transmitter, with an altitude error of about \SI{1.8}{\metre} RMSE
reported in simulation~\cite{gurung-srs-oran-2026}. The downlink arm keeps that
testbed's ray-traced scene, tracker and null discipline and moves the receiver
to the user equipment. What this paper adds is the matched comparison and the
null discipline.
Section~\ref{sec:multipci} carries the second-transmitter route of
\cite{gurung-srs-oran-2026} over to the downlink, where the additional
transmitters are the co-channel cells already licensed at the site.

Section~\ref{sec:metrics} states which 3GPP sensing quantity each quantity
reported here corresponds to. Two of them --- the missed-detection probability
and the false-alarm probability of the second kind --- follow the 3GPP sensing
agreements and are computed in the same association pass as the detection
probability, so they cannot use a different gate from it. The track coverage,
the coverage-to-null ratio and the identity quantities are in-house and have no
3GPP equivalent; they are defined in Section~\ref{sec:metrics}.

Four words recur and are used in one sense throughout. An \emph{arm} is one
transmit--receive configuration scored end to end --- here the LTE-CRS downlink
and the LTE-SRS uplink. A \emph{lap} is one traversal of a road segment by the
target, the unit a coverage figure is computed over. A \emph{cell} is one
simulation run of one arm at one noise seed, the unit $n$ counts. A
\emph{null} is a control run or re-score in which the quantity under test is
removed --- the target, the time order of the ground truth, or its position ---
and whose score is what the arm earns without the thing it claims to measure.

\section{System model}
\label{sec:sysmodel}
The transmitter and the sensing receiver are at fixed, distinct positions
separated by a baseline of length $L$. A point target at position $\mathbf{p}$
is observed through the path transmitter--target--receiver. Writing $R_T$ and
$R_R$ for the transmitter--target and target--receiver ranges, the quantity the
detector resolves in delay is the excess bistatic range
\begin{equation}
  \label{eq:bistatic-range}
  \rho(\mathbf{p}) = R_T + R_R - L ,
\end{equation}
which is zero on the baseline and constant on a prolate spheroid with the two
nodes at its foci. A single transmitter--receiver pair therefore constrains the
target to such a surface and does not by itself observe elevation; the scene's
target height is supplied as a prior.

For a target moving with velocity $\mathbf{v}$, the Doppler shift observed on
that path is
\begin{equation}
  \label{eq:bistatic-doppler}
  f_D = -\frac{1}{\lambda}\,\frac{\mathrm{d}\rho}{\mathrm{d}t}
      = \frac{1}{\lambda}\,\mathbf{v}\cdot(\hat{\mathbf{u}}_T + \hat{\mathbf{u}}_R),
\end{equation}
with $\lambda$ the carrier wavelength, $\hat{\mathbf{u}}_T$ and
$\hat{\mathbf{u}}_R$ the unit vectors from the target toward the transmitter and
the receiver, and the time dependence $\exp(+\mathrm{j}2\pi f t)$, so that a
closing target has $\mathrm{d}\rho/\mathrm{d}t < 0$ and $f_D > 0$. The radial
velocity reported by the detector and used by the association rule and the
tracker is $v_r = -\mathrm{d}\rho/\mathrm{d}t = \lambda f_D$, positive for a
closing target, and $|v_r| \le 2|\mathbf{v}|$ for fixed nodes, with equality
on the baseline extension. This is the bistatic radial-velocity convention
adopted for the 3GPP 6G ISAC evaluations~\cite{ran1-126bis-notes}, where it
is written $2|\mathbf{v}|\cos(\theta_B/2)\cos\delta$ with $\theta_B$ the
bistatic angle and $\delta$ the angle between the velocity and the bisector
(the symbols $\alpha$, $\beta$, $\gamma$ are reserved below for the
TR~38.901 orientation triple). Both \eqref{eq:bistatic-range} and \eqref{eq:bistatic-doppler} are
invariant under exchanging the two nodes, so they take the same form on the
downlink and the uplink arm; what the exchange does change is the angle the
receiving array measures and the position of the second focus.

\subsection{Coordinate and angle conventions}
\label{sec:conventions}
Scene coordinates are a local east--north--up frame with its origin at the
tile's geodetic origin: $x$ is east, $y$ is north, $z$ is up. Azimuth $\phi$ is
measured counter-clockwise from $+x$ and elevation from the horizontal plane, so
a compass bearing $\psi$, measured clockwise from north, enters as
$\phi = 90^\circ - \psi$. Licensed sector azimuths are quoted below as compass
bearings and are converted by that relation. Array orientations are the
bearing--downtilt--slant triple $(\alpha,\beta,\gamma)$ of the global-to-local
rotation of 3GPP TR~38.901~\S7.1.3~\cite{3gpp-tr38901}; that document's zenith angle
is $\theta = 90^\circ - \mathrm{el}$ in the elevation used here.

The sensing receiver is a two-element uniform linear array whose boresight is
its \emph{broadside}, so at $\alpha = 0$ the broadside points east and the
element axis lies north--south. A linear array observes the direction cosine
along its own axis, so the angle it reports is a cone angle $\psi_c$ about that
axis,
\begin{equation}
  \label{eq:cone-angle}
  \sin\psi_c = \sin(\phi - \alpha)\,\cos(\mathrm{el}),
\end{equation}
and that cone is symmetric about the axis: the two half-spaces give the same
measurement and one linear array cannot resolve between them. Every reported
bearing in this work is $\psi_c$ and not a scene azimuth.

\begin{figure}[t]
\centering
\includegraphics[width=0.78\columnwidth]{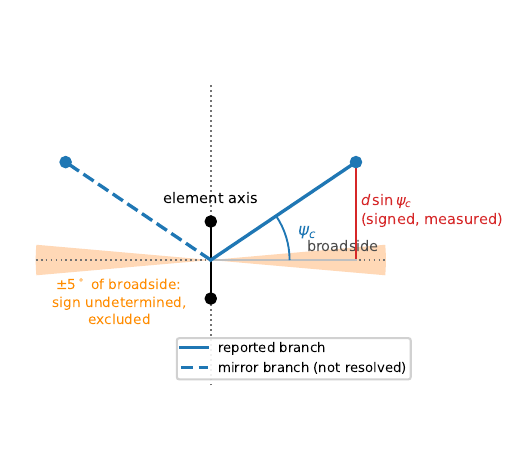}
\caption{What the two-element array measures. The signed along-axis component
$d\sin\psi_c$ is measured, so the sign of the cone angle $\psi_c$ about the
element axis is a measured quantity away from broadside; the front-back branch
about the axis is not resolved by one linear array. Detections within
\ang{5} of broadside, shaded, are excluded from the sign audit because the
along-axis component passes through zero there.}
\label{fig:cone}
\end{figure}

Two distinct things follow, and Fig.~\ref{fig:cone} separates them. The array
measures the \emph{signed} component of the target direction along its own
axis, $d\sin\psi_c$, so the sign of $\psi_c$ is a measured quantity wherever
that component is large. What the single array cannot resolve is the
\emph{front-back} branch about the axis: the two half-spaces that share one
value of $\psi_c$. The first is a convention to be checked, implemented as a single
parameter in the tracker (\texttt{DL\_CONE\_SIGN}, one value per arm, applied where
detections become track candidates and in the filter's measurement model); the
second is a geometric ambiguity no convention removes, and both branches can be
seeded (the birth-branch policy of Section~\ref{sec:siting}). The sign of $\psi_c$ is
checked per arm against the ground-truth cone angle on the detections the
scorer associates.

The target is a single ground vehicle travelling at
\SI{10}{\kilo\metre\per\hour} on a straight trajectory at a height of
\SI{1.6}{\metre}, modelled as a sphere of \SI{1.5}{\metre} radius. The
transmitter and the sensing receiver are both at fixed positions on either side
of that trajectory.

\textbf{Channel.} The channel is rendered offline by a
differentiable ray tracer~\cite{sionna,sionna-rt} on that library's bundled
street-canyon example scene with the building meshes removed, leaving the ground
plane, so every path is either line-of-sight or a ground reflection; the solver
resolves up to six interactions with specular reflection enabled and diffraction
and refraction disabled. Those solver settings are the renderer's defaults and
are not recorded in the trace metadata, which records the scene, the carrier,
the bandwidth, the array geometry and the trajectory. Each arm replays its own
render of the same scene geometry at its own carrier, one channel snapshot per
coherent processing interval, so the channel is held fixed inside each dwell
and the phase progression that a target moving within the dwell would impose
on the burst sequence is not represented; the Doppler the detector measures
is that of the snapshot-to-snapshot range change. The vehicle is the only
moving scatterer.

\section{Arms and their alignment}
\label{sec:arms}

\begin{table}[t]
\caption{The two arms. The bandwidth difference is stated and is not equalised.}
\label{tab:arms}
\centering
\begin{tabular}{@{}lll@{}}
\toprule
 & Arm A & Arm B \\
\midrule
direction        & downlink            & uplink \\
reference signal & LTE CRS             & LTE SRS \\
band             & 3 (FDD)             & 3 (FDD) \\
cell grid, \SI{15}{\kilo\hertz} SCS & 100 PRB & 100 PRB \\
reference-signal span & 100 PRB & 96 PRB \\
occupied BW      & \SI{18.00}{\mega\hertz}        & \SI{17.28}{\mega\hertz} \\
receiver         & UE-side, passive    & eNB-side \\
CPI              & \BthreeCrsCpiS~s    & \BthreeSrsCpiS~s \\
receiver gain    & \BthreeCrsGain      & \BthreeSrsGain \\
CFAR trim        & \BthreeCrsTrim~dB   & \BthreeSrsTrim~dB \\
\bottomrule
\end{tabular}
\end{table}

The following items are equalised between the two arms. Coherent dwell:
\BthreeCrsCpiS~s on arm~A and \BthreeSrsCpiS~s on arm~B, matched to the channel trace's snapshot period so that
one dwell advances one snapshot. Declared false-alarm probability: $10^{-6}$ per
resolution cell on both, with the measured rate quoted against each detector's
own evaluated-cell readback, each bounded by eleven target-absent replays
(Section~\ref{sec:limits}). Association gate: one range grid bin,
\SI{\BthreeCrsGridDeltaRM}{\metre}, and one Doppler bin,
\SI{\BthreeCrsDopGateV}{\metre\per\second} on arm~A and
\SI{\BthreeSrsDopGateV}{\metre\per\second} on arm~B ($\lambda/T_\mathrm{CPI}$),
each read back from the arm's own cells. The range grid
bin is the \SI{\BthreeCrsGridFsMHz}{\mega\hertz}
sample spacing of \SI{\BthreeCrsGridSampleM}{\metre} interpolated
\BthreeCrsGridPad{} times by zero padding of the
\BthreeCrsGridRfft-point transform, while the nominal range resolution set by
the occupied pilot span of \SI{\BthreeCrsPilotApWindowedBwMHz}{\mega\hertz}
is $c/B = \SI{\BthreeCrsPilotApCoverBM}{\metre}$, before the range taper
(\BthreeCrsGridRangeTaper) widens it. Track-to-truth gate: \SI{20}{\metre}, identical for both arms and for
every null. Seed count: $n = \BthreeCrsCovN$ noise realisations per arm, drawn
from the same set. Scorer: one program, one version, invoked identically.
Receiving array element count: two on both arms.

The following items are not equalised.

\textbf{Carrier.} Band~3 is frequency-division duplex, so the downlink and
uplink legs are at \SI{1845}{\mega\hertz} and \SI{1750}{\mega\hertz}, a
\SI{5.4}{\percent} difference. One emulated carrier serves one leg, so the two
arms cannot share a cell and are run separately.

\textbf{Occupied bandwidth.} The sounding reference signal occupies 96 of the
cell's 100 resource blocks, \SI{17.28}{\mega\hertz} against the
\SI{18.00}{\mega\hertz} of arm~A, a difference of \SI{0.72}{\mega\hertz}. The
detector's range cell is set by its range-transform aperture rather than by the
occupied span, and is \BthreeCrsRangeCell~m on arm~A and
\BthreeSrsRangeCell~m on arm~B.

\textbf{Receiver gain treatment.} Each arm is held at a fixed operating point
for the duration of a run, and the two points are not the same number; both are
stated in Table~\ref{tab:arms}. No absolute detection level is claimed from either arm.

\paragraph{Coverage denominator} Coverage is scored over the run length. A
first-light abort makes run length outcome-dependent, so the two arms are
equalised on CPI count before their coverages are compared, or the comparison
is not made.

\paragraph{False-alarm denominator} $P_\mathrm{FA}$ is quoted against the
detector's own evaluated-cell readback; the map size is not used. The two arms do
not evaluate the same number of cells per CPI and the ratio is stated.

\subsection{Processing chain}
\label{sec:chain-proc}
Each arm's receiver extracts the reference-signal resource elements of every
burst it processes, forms a per-burst channel estimate on the pilot grid, and
transforms it to delay with an \BthreeCrsGridRfft-point transform (a
\BthreeCrsGridRangeTaper{} taper over the declared span,
\BthreeCrsGridPad{}-fold zero padding, so one grid bin is
\SI{\BthreeCrsGridDeltaRM}{\metre}), after subtraction of the direct path
and the static taps by an extensive-cancellation step of the stated rank and
a clutter filter. \BthreeCrsGridNDop{} consecutive bursts, one every
\num{20} slots, are transformed to Doppler with a
\BthreeCrsGridDopplerTaper{} taper, giving one range--Doppler map per
coherent interval with a velocity bin of
\SI{\BthreeCrsGridDeltaVMs}{\metre\per\second}. An ordered-statistic CFAR
at the declared false-alarm probability, with the threshold trim of
Table~\ref{tab:arms}, selects candidates inside the direct-path mask and
the range gate; the two receive channels give a cone angle about the array
axis per candidate. Candidates are published as level-C records; the tracker
(an extended Kalman filter in excess bistatic range, range rate and cone
angle, with the target height as a prior) seeds a track on a candidate,
confirms it after three associated updates, and publishes level-D records
that the scorer reads. The scorer associates one reported object with one
target inside one range grid bin and one Doppler bin, and scores coverage on
confirmed tracks within a \SI{20}{\metre} gate against the rendered truth.
Every number quoted for this chain is read back from the cell that produced
it.

\section{Metrics and their 3GPP correspondence}
\label{sec:metrics}

The detection metrics in this paper are the evaluation quantities of the 3GPP
sensing agreements. They are stated in full below, because two of the three
depend on an association rule the specification leaves to the implementer.

\newcommand{\BthreeGateV}{\BthreeSrsDopGateV}

\subsection{Metric and KPI definitions}
\label{sec:metricdefs}

The metrics below are those of 3GPP TR~38.765~\cite{3gpp-tr38765},
clause~4.1, which defines the missed-detection probability and two false-alarm
probabilities over simulation drops and requires a one-to-one association
between detected objects and true targets without fixing the association
method. They are restated here because the same words carry different meanings
in the radar and the cellular literature.

\subsubsection{Evaluation instance}
An evaluation instance is one coherent processing interval of one drop, and
it takes the place of the drop in the definitions of \cite{3gpp-tr38765}: a
drop with $N$ intervals contributes $N$ instances, and all metrics are
accumulated over instances. For instance $n$, let $M_n$
be the number of targets present and $D_n$ the number of those targets not
reported; let $M'_n$ be the number of objects reported and $D'_n$ the number of
reported objects that are not associated with a target. An interval in which the
target is occluded and no echo is rendered contributes to neither sum.

\subsubsection{Detection metrics}
The missed-detection probability is accumulated over instances with at least one
target present, and the type-2 false-alarm probability over instances with at
least one reported object:
\begin{equation}
\mathrm{MDP} = \frac{\sum_n D_n}{\sum_n M_n},
\qquad
\mathrm{FAP_2} = \frac{\sum_n D'_n}{\sum_n M'_n}.
\label{eq:mdpfap}
\end{equation}
With one target per instance, $M_n = 1$, the first expression equals the
report's $P_{\mathrm{md}}$, a mean over drops of $D_n/M_n$. The second is a
ratio of sums over instances, whereas the report's $P_{\mathrm{f2}}$ is the
mean over instances with at least one reported object of $D'_n/M'_n$; the two
differ when the number of reported objects varies between instances, and the
report's form is given beside this one where it is quoted; on the round of
Table~\ref{tab:det} at most \BthreeCrsPfTwoMaxObjects{} object is reported per
instance, so the two forms coincide there (arm~A \BthreeCrsPfTwoCellMean{}
against \BthreeCrsPfTwoReport{} over \BthreeCrsPfTwoNInstances{} instances).
The report's type-1 false-alarm probability, the fraction of target-absent
drops in which any object is reported, is the quantity the target-absent cells
of this paper measure at the reported-object level (Section~\ref{sec:det}).
Both are defined per sensing mode and reported with the association rule that
produced them. Neither definition contains a detection threshold, so
a constant-false-alarm-rate setting is part of the configuration being evaluated
and not a property of the metric.

\subsubsection{Probability of detection and its relation to the missed-detection
probability}
\label{sec:pd}
The radar literature, and the operational specifications a sensing product is
procured against, state detection performance as a probability of detection at a
stated false-alarm rate. The probability of detection used here is the fraction
of observable intervals in which the target is reported and associated:
\begin{equation}
P_D = \frac{1}{N_{\mathrm{obs}}}\sum_{n \in \mathcal{O}} \mathbf{1}\!\left[
\text{target associated in instance } n \right],
\label{eq:pd}
\end{equation}
where $\mathcal{O}$ is the set of observable instances --- those in which the
target is present and an echo is rendered --- and $N_{\mathrm{obs}} =
|\mathcal{O}|$. With one target per instance, $M_n = 1$ for every $n \in
\mathcal{O}$, so \eqref{eq:mdpfap} and \eqref{eq:pd} are the same measurement
read in opposite directions:
\begin{equation}
P_D = 1 - \mathrm{MDP},
\label{eq:pdmdp}
\end{equation}
provided that every instance carries one target, that both figures use the same
observable set $\mathcal{O}$, and that both use the same association rule.

The identity holds only under those three conditions. The denominator must be the observable
instances the receiver actually produced, not the intervals present in the
trajectory: a replay that yields fewer intervals than the trajectory holds
inflates $\mathrm{MDP}$ and deflates $P_D$ by the same ratio. The association
rule must be the one stated in Section~\ref{sec:metrics}; a detection counted
as a hit under a wider gate is a hit for $P_D$ and for $\mathrm{MDP}$ alike, and
the two figures diverge whenever they are computed at different gates. Instances
in which the target is unobservable --- occluded, or inside the direct-path mask
--- must be excluded from both or included in both.

\subsubsection{Association}
$D_n$, $D'_n$ and the indicator in \eqref{eq:pd} are all functions of an
association rule. A reported object is associated with a target when it is that
target's nearest neighbour in normalised range and radial velocity and lies
inside a gate of one range grid bin (\BthreeCrsRangeCell~m on both arms) and one
Doppler bin ($\lambda/T_\mathrm{CPI}$: \BthreeCrsDopGateV~m/s on arm~A,
\BthreeSrsDopGateV~m/s on arm~B), one object to one target.
The specification does not fix the rule; the gate is stated with every result,
and for a fixed set of reports widening it moves $\mathrm{MDP}$ down and
$\mathrm{FAP_2}$ down, since a report that becomes associated is no longer
counted as false.

\subsubsection{Accuracy}
Accuracy is the 90th percentile of the absolute error over associated
detections, reported per quantity.

\subsubsection{Reporting levels}
Measurements are reported at level~C of \cite{3gpp-tr38765}, clause~5.1
(option~C1), one record per interval, carrying excess delay, Doppler, angle,
received power and a confidence value; objects are reported at level~D
(option~D1), one record per associated object, carrying position, radial
velocity and a confidence value. Level~A raw-measurement reporting is not
emitted. The distinction matters for a passive receiver: level~C is observable
with a one-dimensional array, and level~D position is observable only under the
height prior of Section~\ref{sec:track}.

\subsubsection{Target values}
The target levels used as the comparison scale in Table~\ref{tab:det} are
the performance objectives of TR~38.765~\cite{3gpp-tr38765}, Table~4.2-1:
missed-detection probability 5\%, false-alarm probability 5\% for both types,
horizontal and vertical position accuracy 10~m and velocity accuracy 5~m/s,
each at a 90\% confidence level, which the report defines as the 90th
percentile of the error distribution; they were adopted by RAN1 at its
meeting~122bis~\cite{ran1-122bis-notes} and are unbracketed in the approved
report. The report adopts them for the UAV use
case with gNB-based monostatic sensing at 4 or \SI{4.9}{\giga\hertz} (its
Table~6.2-1), so they are a reference scale for a downlink passive bistatic arm
on a ground-vehicle scene, with no matched requirement; every comparison
against them in this paper is labelled as such. The SA1 service KPIs for UAV
intrusion detection, TR~22.837~\cite{3gpp-tr22837} Table~7.2-1, differ (false
alarm 2\%, 10~m at 95\% confidence) and are not used as the scale.

\subsection{In-house quantities}
\label{sec:inhouse}

\begin{table}[t]
\caption{In-house metrics, their definitions, and their nearest 3GPP quantity.}
\label{tab:metrics}
\centering
\begin{tabular}{p{1.5cm}p{2.1cm}p{3.4cm}}
\toprule
ours & nearest 3GPP & difference to state \\
\midrule
$P_D$ (moving) & detection probability, \eqref{eq:pd} & same definition; gated to the moving transit, gate stated \\
MDP & missed-detection probability, \eqref{eq:mdpfap} & $1-P_D$ by construction (Section~\ref{sec:limits}) \\
$\mathrm{FAP_2}$ & type-2 false-alarm probability, \eqref{eq:mdpfap} & the report takes a mean of per-instance ratios and this paper a ratio of sums; identical here because at most \BthreeCrsPfTwoMaxObjects{} object is reported per instance (Section~\ref{sec:det}) \\
$P_\mathrm{f1}$ & type-1 false-alarm probability, TR~38.765 cl.~4.1 & measured on the round's target-absent cell at the reported-object level, per coherent interval rather than per drop \\
coverage & --- & fraction of truth frames with a matched track within the \SI{20}{\metre} scorer gate; denominator is run length \\
permutation null & --- & coverage re-scored against the same truth in permuted time order; what a track that follows the geometry rather than the target earns \\
margin & --- & coverage less the arm's own permutation null, in percentage points; \SI{20}{pp} is the readability gate used here \\
decoy null, margin & --- & coverage against truth displaced 40~m per axis, worst of four; margin is coverage less that worst value, in pp \\
both nulls cleared & --- & coverage exceeds both the permutation null and the worst decoy; the scorer's tracking flag \\
IDF1~\cite{ristani-idf1-2016} & --- & identification F1: harmonic mean of identity precision and recall over truth frames, 1 when one track holds one identity for the whole transit \\
IDSW & --- & identity switches: number of times the track identity assigned to the target changes \\
OSPA / GOSPA~\cite{rahmathullah-gospa-2017} & --- & multi-target distance, order $p=2$, cut-off $c = \SI{20}{\metre}$ (the scorer gate), GOSPA $\alpha = 2$; for a single target a cardinality check \\
RMSE & positioning accuracy, TS~38.857 class & bistatic range error rather than position error \\
NIS & --- & normalised innovation squared of the tracker, median over updates; near its degrees of freedom when the filter's covariance is consistent \\
\bottomrule
\end{tabular}
\end{table}

\begin{table}[t]
\caption{Detection metrics against the reference levels
of~\cite{3gpp-tr38765}, $n = \BthreeCrsCovN$ seeds per arm,
association at one range grid bin and one Doppler bin; the target column is
TR~38.765 Table~4.2-1~\cite{3gpp-tr38765}. Both arms are one round on one
detector build, run back to back, the same round as Table~\ref{tab:track}; the target-absent column is a false-alarm control
and carries no $P_D$ (Section~\ref{sec:det}).}
\label{tab:det}
\centering
\setlength{\tabcolsep}{4pt}
\begin{tabular}{lrrrr}
\toprule
 & target & arm A (CRS) & arm B (SRS) & A/B null \\
\midrule
$P_D$           & ---   & \BthreeCrsPd   & \BthreeSrsPd   & --- \\
MDP             & 0.05  & \BthreeCrsMdp  & \BthreeSrsMdp  & --- \\
$\mathrm{FAP_2}$ & 0.05 & \BthreeCrsFap  & \BthreeSrsFap  & --- \\
$P_\mathrm{f1}$ & 0.05 & ---            & ---            & \BthreeCrsPfOne{} / \BthreeSrsPfOne \\
objects reported & ---  & \BthreeCrsNReported & \BthreeSrsNReported & --- \\
unassociated     & ---  & \BthreeCrsNUnassoc  & \BthreeSrsNUnassoc  & --- \\
raw cand./CPI    & ---  & \BthreeCrsRawMean   & \BthreeSrsRawMean   & \BthreeCrsNullRawMean{} / \BthreeSrsNullRawMean \\
\bottomrule
\end{tabular}
\end{table}

Neither arm approaches the reference MDP or false-alarm level.

\section{Results: detection}
\label{sec:det}
The reference target levels quoted in Table~\ref{tab:det} are the
performance objectives of \cite{3gpp-tr38765}, Table~4.2-1, adopted for the
UAV use case with gNB-based monostatic sensing. They are used here as a reference scale for
two passive bistatic arms on a ground-vehicle scene.

Every figure below carries its target-absent null in the same sentence, on first
use, and every $P_D$ carries the gate it was measured with, because the
Doppler window of the gate is derived from the coherent interval.

The two arms run at the fixed operating points stated in Table~\ref{tab:arms};
arm~A's unrepresentative receiver level is described in Section~\ref{sec:arms},
and no absolute detection level in this paper rests on it.

\begin{figure*}[t]
  \centering
  \includegraphics[width=\textwidth]{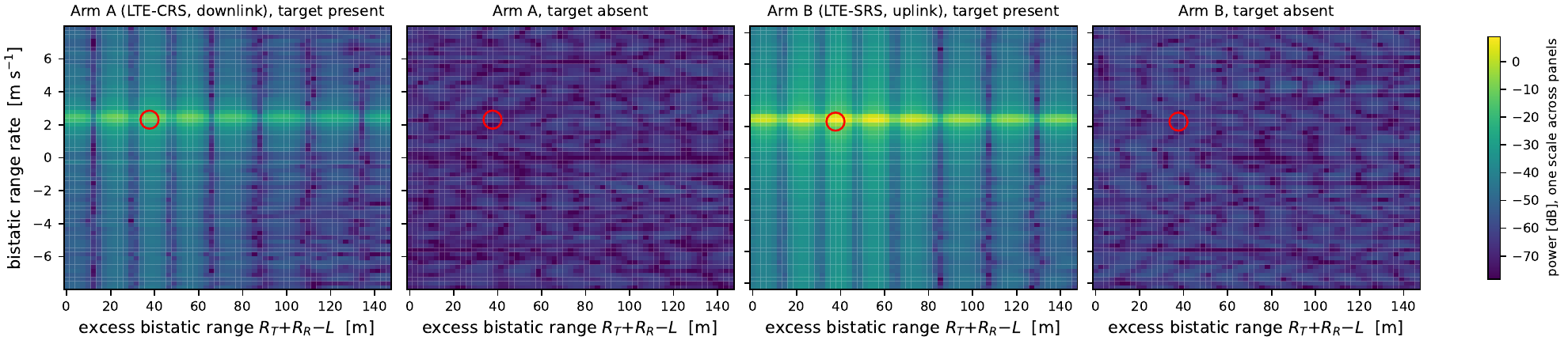}
  \caption{Post-cancellation range--Doppler maps for both arms with the
  target present and with it absent, at one coherent interval (dump index
  \BthreeRdmapCpi), on one colour scale across all four panels. The circle is the
  target's own cell, \BthreeRdmapMarkR~m of excess bistatic range and
  \BthreeRdmapMarkV~m/s of bistatic range rate from the recorded ground truth,
  drawn identically on the target-absent panels, whose render carries no echo.}
  \label{fig:rdmaps}
\end{figure*}

Figure~\ref{fig:rdmaps} shows one coherent interval of each arm's
range--Doppler map beside its target-absent twin. At the target's own cell the
present maps stand \BthreeRdmapCrsPresent~dB (arm~A) and
\BthreeRdmapSrsPresent~dB (arm~B) above their own map median, against
\BthreeRdmapCrsAbsent~dB and \BthreeRdmapSrsAbsent~dB in the absent twins. On
the marked Doppler row the peak lies at \BthreeRdmapPeakR~m, within one
\BthreeCrsRangeCell~m grid bin of the truth, and stays within
\SI{3}{\decibel} of that peak over \BthreeRdmapRowLo--\BthreeRdmapRowHi~m,
which is wider than the resolution cell, so the figure establishes the target's
range at the peak and not the width of the response. The two arms run different
receiver-gain treatments (Section~\ref{sec:arms}), so each panel is read against
its own null and not against the other arm's level.

Arm A (LTE-CRS, downlink, the UE receives) holds
$P_D = \BthreeCrsPd \pm \BthreeCrsPdSd$ over
$n = \BthreeCrsPdN$ seeds on $\BthreeCrsNObsMov$ gated observation
opportunities with a \BthreeCrsRangeCell~m range cell. Arm B (LTE-SRS, uplink,
the eNB receives) holds
$P_D = \BthreeSrsPd \pm \BthreeSrsPdSd$ over
$n = \BthreeSrsPdN$ seeds on $\BthreeSrsNObsMov$ opportunities with
a \BthreeSrsRangeCell~m range cell.

$P_D$ is conditional on an echo being rendered: intervals in which the
target is occluded or outside the gate contribute to neither the numerator
nor the denominator, so the figure is a detection probability given
visibility, and the number of gated observation opportunities beside each
$P_D$ is the denominator that conditioning leaves. On these two arms the
condition removes almost nothing: arm~A processes
\BthreeCrsSchedCpis{} coherent intervals per cell and presents
\BthreeCrsVisibleObs{} moving instances, a removed fraction of
\SI{\BthreeCrsRemovedPct}{\percent}, and arm~B
\BthreeSrsSchedCpis{} against \BthreeSrsVisibleObs{}, a removed fraction of
\SI{\BthreeSrsRemovedPct}{\percent} with a worst cell at
\SI{\BthreeSrsRemovedPctWorst}{\percent}.

The target-absent cell of each arm measures false alarms. The detector's mean
raw candidate count per coherent interval is \BthreeCrsNullRawMean{} on arm~A
over \BthreeCrsNullCpis{} intervals and \BthreeSrsNullRawMean{} on arm~B over
\BthreeSrsNullCpis{}. The type-1 false-alarm probability of TR~38.765
cl.~4.1 is measured on the same cells at the reported-object level: arm~A
publishes \BthreeCrsPfOneObjects{} objects over \BthreeCrsPfOneCpis{}
intervals and arm~B \BthreeSrsPfOneObjects{} over \BthreeSrsPfOneCpis{}, so
$P_\mathrm{f1} = \BthreeCrsPfOne$ and $\BthreeSrsPfOne$ on one cell per
arm, which bounds the estimate at the cell level. Below the object level the
threshold is crossed in \BthreeCrsPfOneRawCpis{} of \BthreeCrsPfOneCpis{}
intervals on arm~A and \BthreeSrsPfOneRawCpis{} of \BthreeSrsPfOneCpis{} on
arm~B.

The arm-to-arm difference is reported per round, paired by noise seed. In
the round of Tables~\ref{tab:det} and~\ref{tab:track} arm~A's $P_D$ exceeds
arm~B's by $\BthreeArmDPdDtwo \pm \BthreeArmDPdDtwoSe$ on
\BthreeArmDPdDtwoPos{} of \BthreeArmDPdDtwoN{} seeds, and in an earlier round
on the same build by $\BthreeArmDPdRep \pm \BthreeArmDPdRepSe$ on
\BthreeArmDPdRepPos{} of \BthreeArmDPdRepN{}; both exceed the execution
threshold of Section~\ref{sec:variance}, so the detection-rate ordering holds
on this build, and its attribution among the four unequalised items of
Section~\ref{sec:arms} is open.

\subsection{Reproduction with a second receiver implementation}
\label{sec:altrx}
Arm~A's detection rate is a property of the waveform, the geometry and the
detector configuration, or it is a property of one receiver implementation. To
separate these, the same scene was run through an independent downlink
cell-specific reference signal receiver, built on a different software radio
stack, at the same coherent dwell, pilot count, burst rate, receiver level,
transit window and detection threshold as arm~A, and scored by the same scorer
at the same gate.

Over $n = \BthreeCrsAltCells$ noise realisations the second receiver reports a
detection probability of $\BthreeCrsAltPd \pm \BthreeCrsAltPdSd$ against
$\BthreeCrsPd \pm \BthreeCrsPdSd$ for the arm~A cells run beside it; the two
agree within their spreads. Its track coverage does not agree: under the
confirmed-track convention it reads $\BthreeCrsAltCov \pm \BthreeCrsAltCovSd$~\%
against $\BthreeCrsCov \pm \BthreeCrsCovSd$~\%, with \BthreeCrsAltPass{} of
\BthreeCrsAltCells{} cells clearing both nulls, because its tracks do not
reach the confirmed state inside the gate. The second receiver
reproduces arm~A's detection rate and bounds the detector; it is a control and
does not appear in the tables.

\section{Results: tracking}
\label{sec:track}

\begin{table}[t]
\caption{Tracking, band 3 vehicle, matched scorer. Null column is the
target-absent arm scored identically.}
\label{tab:track}
\centering
\begin{tabular}{lrrrr}
\toprule
 & Arm A & A null & Arm B & B null \\
\midrule
coverage (\%)     & \BthreeCrsCov      & \BthreeCrsNullCov      & \BthreeSrsCov      & \BthreeSrsNullCov \\
permutation null (\%) & \BthreeCrsCovNull & --- & \BthreeSrsCovNull & --- \\
OSPA (m)          & \BthreeCrsOspa     & \BthreeCrsNullOspa     & \BthreeSrsOspa     & \BthreeSrsNullOspa \\
GOSPA loc (m)     & \BthreeCrsGospaLoc & \BthreeCrsNullGospaLoc & \BthreeSrsGospaLoc & \BthreeSrsNullGospaLoc \\
IDF1              & \BthreeCrsIdf      & \BthreeCrsNullIdf      & \BthreeSrsIdf      & \BthreeSrsNullIdf \\
ID switches       & \BthreeCrsIdsw     & \BthreeCrsNullIdsw     & \BthreeSrsIdsw     & \BthreeSrsNullIdsw \\
RMSE (m)          & \BthreeCrsRmse     & \BNOTRACKS     & \BthreeSrsRmse     & \BNOTRACKS \\
NIS median        & \BthreeCrsNis      & \BthreeCrsNullNis      & \BthreeSrsNis      & \BthreeSrsNullNis \\
CPIs              & \BthreeCrsCpis     & \BthreeCrsNullCpis     & \BthreeSrsCpis     & \BthreeSrsNullCpis \\
\midrule
worst decoy (\%)  & \BthreeCrsDecoyWorst & \BthreeCrsNullDecoyWorst & \BthreeSrsDecoyWorst & \BthreeSrsNullDecoyWorst \\
decoy margin (pp) & \BthreeCrsDecoyMarginPp & --- & \BthreeSrsDecoyMarginPp & --- \\
\bottomrule
\end{tabular}
\end{table}

Coverage counts truth frames with a matched track and says nothing about
how long any one track is held: an arm can cover most of the transit through
a succession of short tracks, which IDF1 and the identity-switch count of
Table~\ref{tab:track} record and coverage does not. The two are reported
together for that reason, and a coverage figure quoted alone is not a
statement of sustained tracking.

Measured on the same records, the same window and the same gate as coverage,
the two arms separate on continuity. Arm~A's covered frames
fall into \BthreeCrsContRunsPerCell{} runs per cell, the longest of them
\BthreeCrsContLongestFrames{} frames
--- \SI{\BthreeCrsContLongestS}{\second} and
\SI{\BthreeCrsContLongestM}{\metre} of the ground truth's own path --- and
arm~B's into \BthreeSrsContRunsPerCell{} runs with a longest run of
\SI{\BthreeSrsContLongestM}{\metre}. Coverage is held in one or two long runs per cell on both arms. A covered
transit of \SI{\BthreeCrsContLongestM}{\metre} held under one identifier for
\SI{\BthreeCrsContIdLongestM}{\metre} is what the IDF1 column of
Table~\ref{tab:track} reports.

\subsection{Null gates for coverage}
\label{sec:nullgates}
\label{sec:twogates}

An absent cell on this testbed has been credited coverage with zero detections,
so a coverage figure is quoted here only beside two nulls, both scored by the same
code on the same cells.

\begin{itemize}
\item The \emph{permutation} null re-scores each arm against a permuted
  ground-truth order. Arm A reaches \BthreeCrsCov~\% against
  \BthreeCrsCovNull~\%, arm B \BthreeSrsCov~\% against \BthreeSrsCovNull~\%.
\item The \emph{decoy} null re-scores each arm against ground truth displaced
  40~m along each axis in turn and keeps the worst. Arm A's worst decoy is
  \BthreeCrsDecoyWorst~\%, arm B's \BthreeSrsDecoyWorst~\%.
\end{itemize}

The margin of coverage over the worst decoy, in percentage points, is
reported with every coverage figure (arm~A \SI{\BthreeCrsDecoyMarginPp}{pp},
arm~B \SI{\BthreeSrsDecoyMarginPp}{pp}); a cell whose coverage does not
exceed both its permutation null and its worst decoy is not credited with
tracking, which is the two-null rule of Table~\ref{tab:metrics} applied
throughout, and the margin over the permutation null states by how much. The
nulls are controls with a known answer, and the pass rule is a comparison
against them with a \SI{5}{pp} margin and is a pass rule, with no significance
level attached. The
coverage-to-decoy ratio of an earlier draft is not reported, since it is
undefined where the worst decoy is zero (\BthreeCrsDecoyRatioUndef{} of arm~A's
\BthreeCrsCells{} cells).

\subsection{Run-to-run variance}
\label{sec:variance}
The noise seed does not define a reproducible experimental unit. The five
seeds of arm~B were re-executed in four further blocks on one build with the
same configuration and the same input trace (checked by hash), 20 cells on an
idle host: the execution standard deviation, pooled within seed, is
\num[round-mode=places,round-precision=3]{\BthreeQuietSrsExecPdPooled} in
$P_D$, \SI[round-mode=places,round-precision=1]{\BthreeQuietSrsExecCovPooled}{pp}
in coverage and \num[round-mode=places,round-precision=3]{\BthreeQuietSrsExecIdfPooled}
in IDF1, against between-seed standard deviations of
\num[round-mode=places,round-precision=3]{\BthreeQuietSrsExecPdSeedPooled},
\SI[round-mode=places,round-precision=1]{\BthreeQuietSrsExecCovSeedPooled}{pp}
and \num[round-mode=places,round-precision=3]{\BthreeQuietSrsExecIdfSeedPooled}.
The number of coherent intervals scored moves on identical inputs,
\BthreeSrsExecQuietCpiMin{} to \BthreeSrsExecQuietCpiMax{} over
\BthreeSrsExecQuietN{} cells (43 in one cell that overlapped other work on the
host), and it is the denominator of coverage and $P_D$; host contention is one
cause and the emulator's scheduling and acquisition time are not separated
from it. Every $\pm$ in this paper is a between-cell standard deviation that
contains execution variance, and an arm-to-arm difference below
\BthreeSrsExecResolvePd{} of $P_D$ or \SI{\BthreeSrsExecResolveCovPp}{pp} of
coverage is not resolvable at the $n$ reported here.

\section{A real site geometry: High St, Reservoir}
\label{sec:highst}
The arms above run on a synthetic street canyon. This section repeats arm~A on a
surveyed Australian site.

\subsection{Scene and geometry}
\begin{figure*}[t]
\centering
\includegraphics[width=\textwidth]{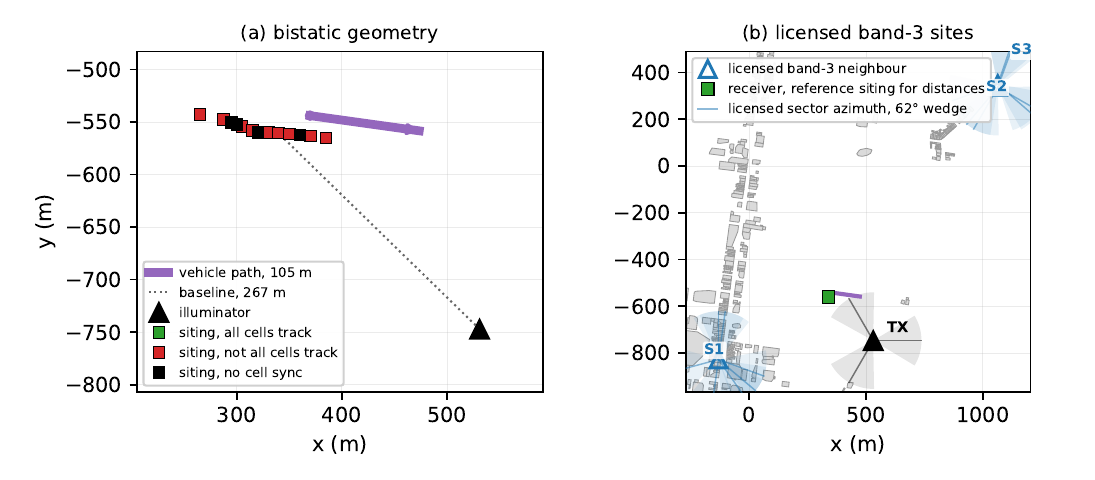}
\caption{High St Reservoir layout, drawn from the scene's own meshes, the rendered
trace, and the ACMA band-3 licence records: (a) the bistatic geometry of the
\SI{\HstLapLen}{\metre} lap and every receiver siting tested, coloured by
outcome; (b) the illuminator and the
\HstNbrN{} nearest licensed band-3 sites with their sector azimuths.
Direct-path levels in Table~\ref{tab:sites} are free-space bounds at the
licensed maximum EIRP, which is the right regime here because the two-ray break
point is \SI{3.3}{\kilo\metre}.}
\label{fig:highst_layout}
\end{figure*}

The scene is built from OpenStreetMap building footprints for a
\SI{1900}{\metre} square centred on High St, Reservoir, Victoria, and rendered
with the same ray tracer and the same two-tap channel as every other cell in
this paper. The transmitter is a band-3 LTE eNB at
$(\HstEnbX, \HstEnbY, \HstEnbZ)$~\si{\metre}; the passive receiver is a
stationary UE at $(\HstUeX, \HstUeY, \HstUeZ)$~\si{\metre}, giving a
\SI{\HstBaseline}{\metre} baseline. The target is a car on a straight
\SI{\HstLapLen}{\metre} section of Pender St, driven one way at
\SI{\HstSpeed}{\metre\per\second} (\SI{\HstSpeedKmh}{\kilo\metre\per\hour}) for
\SI{\HstLapS}{\second}, which the \SI{\HstCpiDt}{\second} coherent interval
samples as \HstSlowOnCpis{} intervals. Carrier \SI{\HstFcMHz}{\mega\hertz},
bandwidth \SI{\HstBwMHz}{\mega\hertz}, as in Table~\ref{tab:arms}.
Table~\ref{tab:highst} states the simulation conditions in full.
Figure~\ref{fig:highst_layout} shows the layout. The nearest building footprint is \SI{178}{\metre} from the lap and \SI{291}{\metre} from the \SI{315}{\metre} receiver siting, so this geometry is open ground rather than a built-up street, and the clutter statements of Section~\ref{sec:limits} are read in that light. The illuminator is licensed site \HstTxSid{}, whose record places it at the simulated transmitter position and height; its licensed carrier is \SI{\HstTxLicFc}{\mega\hertz} in \SI{\HstTxLicBw}{\mega\hertz}, while the render uses \SI{\HstFcMHz}{\mega\hertz} in \SI{\HstBwMHz}{\mega\hertz}: the geometry is taken from the licence record.

Figure~\ref{fig:highst_layout}(b) and Table~\ref{tab:sites} are read as a
deployment statement. The \HstNbrN{} neighbours are
licensed sites, not rendered transmitters: every result here uses one
illuminator, because the emulated receiver does not hold synchronisation when
more than one cell identity is present. In the field that restriction does not
apply, and the route is per-identity extraction on a single recording. Cells on
the same carrier arrive in the same samples, so one capture at one receiver
yields one bistatic pair per co-channel cell, each with its own frame offset and
its own delay origin, and the reference signals of five of every six
neighbouring identities fall on a different subcarrier comb from the
illuminator's. Two of the table's columns decide which
neighbours can contribute at all. A neighbour is a usable pair only if it is
co-channel with the illuminator and one of its sectors faces the receiver, and
those two conditions do not follow distance: the nearest neighbour at
\SI{521}{\metre} is \SI{51}{\degree} off its best sector, the strongest is
boresight-on but carries no co-channel component, and the only neighbour that is
both co-channel and within \SI{30}{\degree} is the furthest at
\SI{1177}{\metre}. The illuminator itself is a three-sector site, and the arm
reported here was rendered on the sector \SI{19}{\degree} off the receiver.

\begin{table*}[t]
\caption{Licensed band-3 sites at the scene, from the ACMA record subset
(\HstSiteRecN{} sector-carrier records over \HstSiteN{} sites). TX is the
simulated illuminator; S1--S3 are the nearest neighbours, with distance measured
from the \SI{315}{\metre} receiver siting.}
\label{tab:sites}
\centering
\small
\begin{tabular}{@{}llrrrrrrrr@{}}
\toprule
& site & $d$ & $h$ & carriers & BW & EIRP\textsuperscript{a} & $P_\mathrm{rx}$\textsuperscript{b} & AZ\textsuperscript{c} & $\Delta\phi$\textsuperscript{d} \\
& & (m) & (m) & (MHz) & (MHz) & (dBW) & (dBm) & (deg) & (deg) \\
\midrule
TX & 102006 & 267 & 29.0 & 1812 & 15 & 32.8 & -23 & 330 & -16 \\
S1 & 303268 & 541 & 15.3 & 1812--1858 (5) & 2.5, 12.5, 15 & 41.4 & -21 & 110 & -50 \\
S2 & 302246 & 1153 & 14.2 & 1826--1858 (4) & 2.5, 12.5, 15 & 41.6 & -27 & 220 & -1 \\
S3 & 152292 & 1163 & 17.0 & 1812 & 15 & 30.8 & -38 & 250 & -31 \\
\bottomrule
\end{tabular}

\vspace{2pt}
\footnotesize
\textsuperscript{a}Licensed maximum EIRP, not a measured transmit power.
\textsuperscript{b}Free-space direct-path level at the receiver at that EIRP, an
upper bound; ground reflection adds at most about \SI{6}{\decibel} at these
ranges.
\textsuperscript{c}The licensed sector azimuth, of that site's several, whose
boresight comes closest to the \SI{315}{\metre} receiver siting.
\textsuperscript{d}The angle off that sector's boresight. A sector pointed away
is not a usable bistatic pair at any distance or EIRP: the nearest neighbour here
is the worst-pointed, and the only co-channel neighbour within
\SI{30}{\degree} is the furthest.
\end{table*}

\begin{table}[t]
\caption{Simulation conditions, High St Reservoir. Geometry and waypoints are read
from the rendered trace's own metadata; detector settings are those of the
CRS arm of Table~\ref{tab:arms}.}
\label{tab:highst}
\centering
\footnotesize
\begin{tabular}{@{}ll@{}}
\toprule
\textbf{Scene} & \\
source & OpenStreetMap footprints, \SI{1900}{\metre} square \\
geometry & 451 building meshes, 7829 triangles \\
buildings & extruded, not flattened \\
propagation & Sionna RT, two static taps \\
\midrule
\textbf{Geometry} & \\
transmitter (eNB) & $(\HstEnbX, \HstEnbY, \HstEnbZ)$ m \\
receiver (UE) & $(\HstUeX, \HstUeY, \HstUeZ)$ m \\
baseline & \SI{\HstBaseline}{\metre} \\
line of sight & unobstructed (mesh ray cast) \\
tx array & $1\times2$ TR~38.901, \ang{0} \\
rx array & $1\times2$ isotropic, \ang{0} \\
\midrule
\textbf{Target} & \\
path & straight, \SI{\HstLapLen}{\metre} on Pender St \\
speed & \SI{\HstSpeed}{\metre\per\second} (\SI{\HstSpeedKmh}{\kilo\metre\per\hour}) \\
duration & \SI{\HstLapS}{\second}, one way \\
model & sphere, \SI{1.5}{\metre} radius, at \SI{1.6}{\metre} \\
mean RCS, path mean & \SI{+18.45}{\decibel} re \SI{1}{\square\metre} \\
\midrule
\textbf{Waveform and detector} & \\
carrier / bandwidth & \SI{\HstFcMHz}{\mega\hertz} / \SI{\HstBwMHz}{\mega\hertz} \\
reference signal & LTE CRS, downlink \\
burst stride / CPI & 20 slots / 64 bursts \\
coherent dwell & \SI{640}{\milli\second} $=\Delta t$ \\
range grid bin / velocity bin & \SI{\BthreeCrsGridDeltaRM}{\metre} / \SI{\BthreeCrsGridDeltaVMs}{\metre\per\second} \\
CFAR & OS, rank 0.75, \SI{8192}{} FFT \\
direct-path mask & \SI{24.4}{\metre} \\
range gate & \SI{85}{\metre} \\
velocity bound & \SI{7.6}{\metre\per\second} \\
clutter knee & \SI{2.031}{\metre\per\second} \\
\midrule
\textbf{Round} & \\
cells per arm & 15 present, 5 absent \\
varied between cells & noise seed only \\
\bottomrule
\multicolumn{2}{@{}p{\columnwidth}@{}}{\vspace{2pt}\scriptsize
Section~\ref{sec:longtraj} reports arms that differ from this configuration in
lap length (\SI{\LtLapM}{\metre}), ground speed
(\SI{\LtLapSpeed}{\metre\per\second}), receiver position (the \SI{315}{\metre}
siting of Section~\ref{sec:siting}, at \SI{10}{\metre}), range gate
(\SI{300}{\metre}), target height and the range taper (occupied band).} \\
\end{tabular}
\end{table}

The lap was sited so that no part of it falls inside the direct-path mask: the
excess bistatic range stays between \SI{27.2}{\metre} and \SI{69.6}{\metre},
clearing the \SI{24.4}{\metre} mask throughout and leaving \SI{15.4}{\metre} of
margin against the \SI{85}{\metre} range gate. The receiver array is a
two-element isotropic pair; the transmitter array follows TR~38.901. Both sit at
\ang{0} boresight.

\subsection{Detection}
Detection on the site is reported at the same operating point as the
synthetic scene. Each cell yields \HstSlowOnNdet{}
moving detections over \HstSlowOnCpis{} coherent intervals at
$n = \HstSlowOnN$, with the delay origin resolved to the same range bin on every
interval of every cell and the direct path standing \SI{44.5}{\decibel} above the
profile median. No observation falls inside the masked band. The
\HstNullN{} target-absent cells, the same renders with the echo zeroed, return
zero moving records.

\subsection{Tracking and null gates}
\label{sec:highsttrack}
Coverage on this geometry is \SI{\HstPoolCov}{\percent} with a standard
deviation of \SI{\HstPoolCovSd}{\percent} over $n = \HstPoolN$ cells that differ
only in noise seed. Read against the two-gate criterion of
Section~\ref{sec:nullgates}, \HstPoolPass{} of \HstPoolN{} cells demonstrate
tracking; those that do have coverage \SI{\HstPoolCovPass}{\percent}.

\subsection{Trajectory length and target height}
\label{sec:longtraj}
\begin{figure*}[t]
\centering
\includegraphics[width=\textwidth]{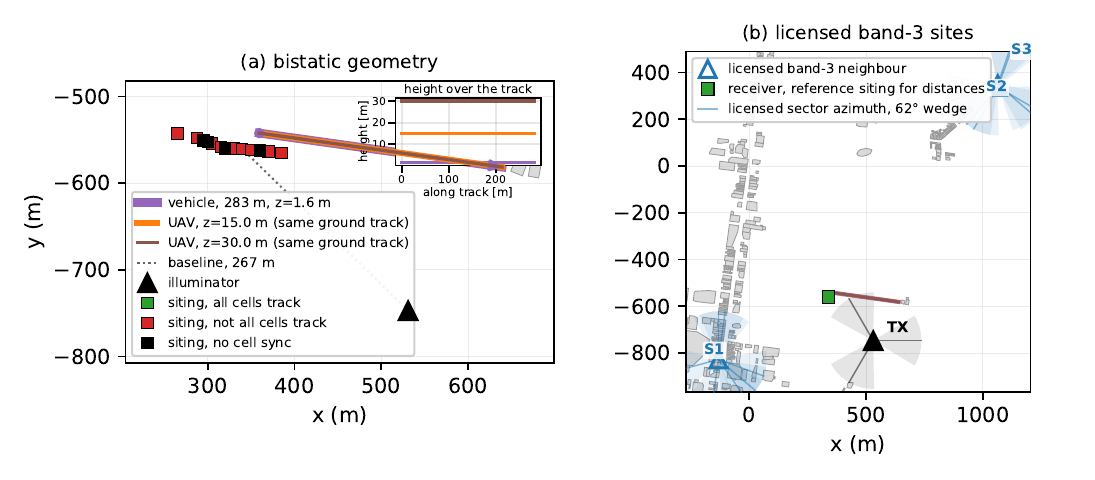}
\caption{The \SI{\LtLapM}{\metre} Pender St lap at target heights
\SI{\LtVehZ}{\metre}, \SI{\LtUavAZ}{\metre} and \SI{\LtUavBZ}{\metre} over one
common ground track, with the heights inset: (a) the bistatic geometry, with the
receiver-siting markers of the \SI{\HstLapLen}{\metre} round
(Section~\ref{sec:siting}), which are not re-evaluated on this lap; (b) the
illuminator and the \HstNbrN{} nearest licensed band-3 sites with their sector
azimuths. Direct-path levels and sector angles are those of Table~\ref{tab:sites}.}
\label{fig:longtraj_layout}
\end{figure*}

The lap is extended to \SI{\LtLapM}{\metre} at
\SI{\LtLapSpeed}{\metre\per\second} and repeated with the target at
\SI{\LtUavAZ}{\metre} and \SI{\LtUavBZ}{\metre} above ground over the same
ground track, so target height is the only quantity that differs between the
three arms. Figure~\ref{fig:longtraj_layout} draws the three trajectories
over the layout of Figure~\ref{fig:highst_layout}(a). The receiver is at the
\SI{315}{\metre} siting of Section~\ref{sec:siting}, raised to \SI{10}{\metre}, and
the range gate is \SI{300}{\metre}; Table~\ref{tab:highst} notes every other
difference from the \SI{\HstLapLen}{\metre} round.
Table~\ref{tab:longtraj} reports each arm over cells that differ only in noise
seed, with the detector's range taper applied over the occupied subcarrier band
rather than over the full transform grid; every other figure in this paper was
measured with the full-grid taper, which on this lap gives detection rates of
\LtZVehPd{} $\pm$ \LtZVehPdSd{} ($n = \LtZVehN$), \LtZUavAPd{} $\pm$
\LtZUavAPdSd{} ($n = \LtZUavAN$) and \LtZUavBPd{} $\pm$ \LtZUavBPdSd{}
($n = \LtZUavBN$) for the same three arms, identity scores of \LtZVehIDF{},
\LtZUavAIDF{} and \LtZUavBIDF{}, and the same null. The occupied-band taper
raises the detection rate by \SI{\TaGainLoPct}{\percent} to
\SI{\TaGainHiPct}{\percent} on these arms; its false-alarm rate on
target-absent cells is \num{\TaPfaOn} per scanned cell against \num{\TaPfaOff}
with the full-grid taper.

Detection and identity do not order the same way. The airborne arm at
\SI{\LtUavAZ}{\metre} has the highest detection rate, \LtUavAPd{} $\pm$
\LtUavAPdSd{}, and the lowest identity score, \LtUavAIDF{} $\pm$
\LtUavAIDFSd{}; the ground arm detects least often, \LtVehPd{} $\pm$
\LtVehPdSd{}, and holds one identity on \LtVehIdswZero{} of \LtVehN{} cells.
The arm at \SI{\LtUavBZ}{\metre} sits between them on identity with fewer
switches than the lower one. On this geometry the airborne arms exceed the
ground arm in detection rate and fall below it in identity continuity.

\begin{figure*}[t]
\centering
\includegraphics[width=\textwidth]{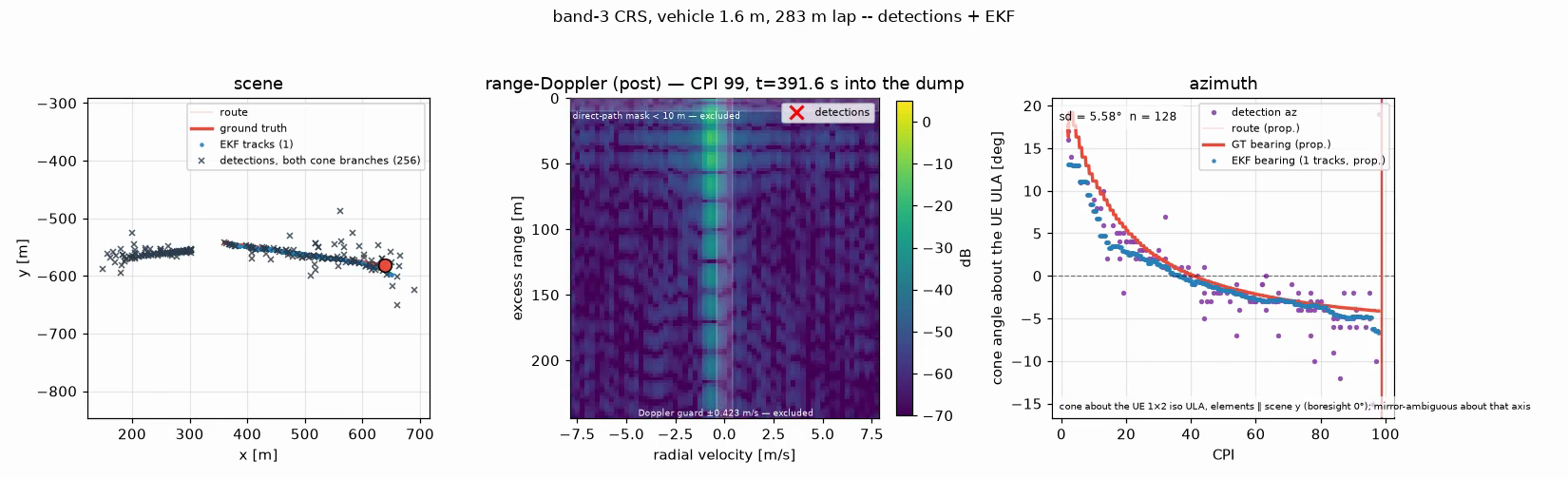}\\[2pt]
\includegraphics[width=\textwidth]{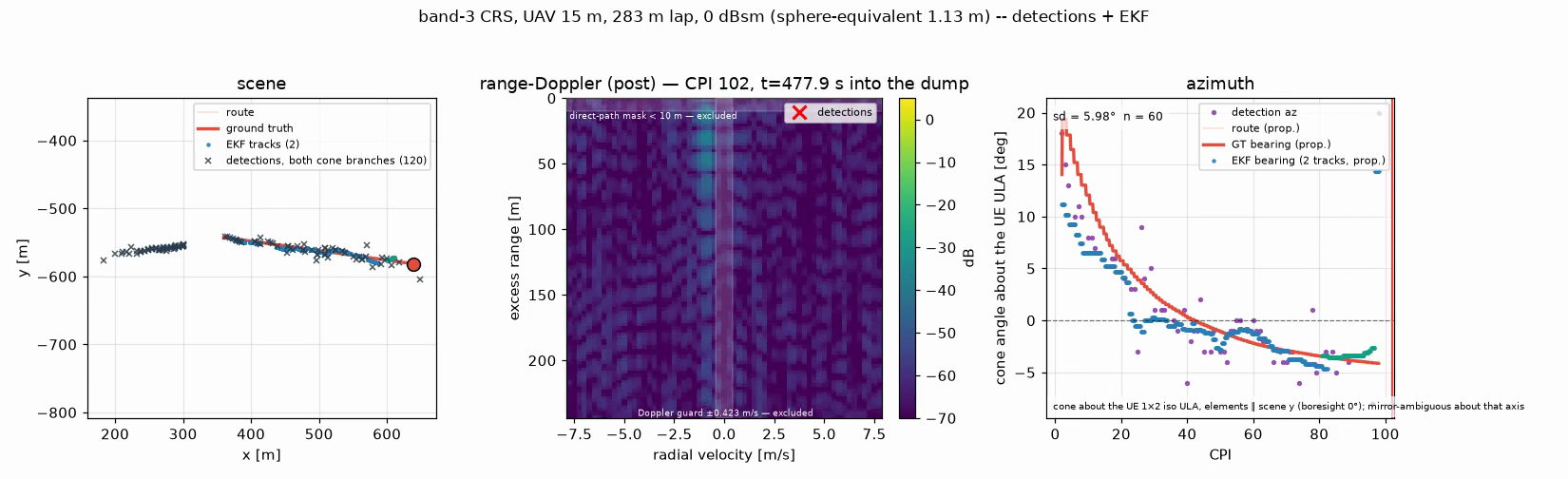}
\caption{Final frames of the animations of the \SI{\LtLapM}{\metre} lap, one
cell each: top, the vehicle at \SI{\LtVehZ}{\metre}; bottom, the airborne target
at \SI{\LtUavAZ}{\metre} rendered at a mean bistatic RCS of \SI{0}{\decibel} re
\SI{1}{\square\metre} (\SI{1.13}{\metre} sphere equivalent), smaller than the
target of Table~\ref{tab:longtraj}. Panels: the scene with ground truth,
detections on both branches of the receive cone and the tracker output; the
post-cancellation range--Doppler map of the last coherent interval with the
direct-path mask and Doppler guard marked; the detected and tracked cone angle
against ground truth per interval.}
\label{fig:demo_frames}
\end{figure*}

Figure~\ref{fig:demo_frames} shows the final frame of the animation of one cell
of the vehicle arm and of one cell with the airborne target at
\SI{\LtUavAZ}{\metre} rendered at \SI{0}{\decibel} re \SI{1}{\square\metre},
the lowest rung of a \SI{0}{}, \SI{-10}{} and \SI{-20}{\decibel} re
\SI{1}{\square\metre} ladder at which the arm still tracks; the ladder's
detection rates are held in the repository record.

\begin{table}[t]
\caption{The \SI{\LtLapM}{\metre} Pender St lap at three target heights with the
range taper over the occupied band, cell scorer at a \SI{20}{\metre} gate, cells
differing only in noise seed, $\pm$ one standard error of the mean. Each arm's
target-absent re-render covers \SI{\LtVehNullCov}{\percent}.}
\label{tab:longtraj}
\centering
\footnotesize
\setlength{\tabcolsep}{2.6pt}
\begin{tabular}{lcccccc}
\toprule
target & $z$ & $n$ & $P_\mathrm{D}$ & IDF1 & IDSW & GOSPA \\
 & [m] & & & & & [m] \\
\midrule
vehicle & \LtVehZ  & \LtVehN  & $\LtVehPd  \pm \LtVehPdSd$  & $\LtVehIDF  \pm \LtVehIDFSd$  & \LtVehIdsw  & $\LtVehGospa  \pm \LtVehGospaSd$  \\
UAV     & \LtUavAZ & \LtUavAN & $\LtUavAPd \pm \LtUavAPdSd$ & $\LtUavAIDF \pm \LtUavAIDFSd$ & \LtUavAIdsw & $\LtUavAGospa \pm \LtUavAGospaSd$ \\
UAV     & \LtUavBZ & \LtUavBN & $\LtUavBPd \pm \LtUavBPdSd$ & $\LtUavBIDF \pm \LtUavBIDFSd$ & \LtUavBIdsw & $\LtUavBGospa \pm \LtUavBGospaSd$ \\
\bottomrule
\end{tabular}
\end{table}

\subsection{Receiver siting}
\label{sec:siting}
Over the whole lap the target subtends \ang{\HstSitBSub} of azimuth at the
reported receiver siting, so the measurement is close to range-only and a
decoy displaced along a locus of constant bistatic range is close to
indistinguishable from the target. Whether a receiver placed elsewhere on the
kerbside line, where the lap subtends more, tracks better is the question this
subsection measures. Nine positions along that line were run with the
transmitter, the lap, the image and the operating point held fixed, at
$n = \HstSitAN$ noise seeds per position, on one detector build and one
scoring convention, with one birth branch seeded (the policy question is
returned to below). Table~\ref{tab:siting} gives the subtense, the coverage and
the pass fraction at each position.

\begin{table}[t]
\caption{Receiver siting along the street, one birth branch, $n = \HstSitAN$
seeds per position on one build and one scoring convention. Coverage is the
mean over seeds with one sample standard deviation; pass is the number of cells
clearing both nulls. The lap's angular subtense at the receiver does not order
the coverage.}
\label{tab:siting}
\centering
\small
\begin{tabular}{@{}rrr@{}}
\toprule
subtense (deg) & coverage (\%) & pass \\
\midrule
\HstSitASub & \HstSitACov\,$\pm$\,\HstSitACovSd & \HstSitAPass/\HstSitAN \\
\HstSitBSub & \HstSitBCov\,$\pm$\,\HstSitBCovSd & \HstSitBPass/\HstSitBN \\
\HstSitCSub & \HstSitCCov\,$\pm$\,\HstSitCCovSd & \HstSitCPass/\HstSitCN \\
\HstSitDSub & \HstSitDCov\,$\pm$\,\HstSitDCovSd & \HstSitDPass/\HstSitDN \\
\HstSitESub & \HstSitECov\,$\pm$\,\HstSitECovSd & \HstSitEPass/\HstSitEN \\
\HstSitFSub & \HstSitFCov\,$\pm$\,\HstSitFCovSd & \HstSitFPass/\HstSitFN \\
\HstSitGSub & \HstSitGCov\,$\pm$\,\HstSitGCovSd & \HstSitGPass/\HstSitGN \\
\HstSitHSub & \HstSitHCov\,$\pm$\,\HstSitHCovSd & \HstSitHPass/\HstSitHN \\
\HstSitISub & \HstSitICov\,$\pm$\,\HstSitICovSd & \HstSitIPass/\HstSitIN \\
\bottomrule
\end{tabular}

\vspace{2pt}
\footnotesize
Both-branch policy, run at the two extreme sitings only:
\ang{\HstSitASub} reads \HstSitABrCov\,$\pm$\,\HstSitABrCovSd~\% with
\HstSitABrPass/\HstSitABrN{} passing, and \ang{\HstSitISub} reads
\HstSitIBrCov\,$\pm$\,\HstSitIBrCovSd~\% with \HstSitIBrPass/\HstSitIBrN.
Coverage is lower in both and the spread is larger; the pass fraction at
\ang{\HstSitASub} is the one quantity the second branch improves.
\end{table}

\textbf{Birth-branch policy.} The relay forms a track seed from a detection's
range, range rate and cone angle; the cone is ambiguous about the array axis,
so a seed can be placed on either of two mirror solutions, and seeding both
doubles the seed count and the association load. The two policies were run at
the two ends of the line only (table footnote). Seeding both lowers the mean
coverage and widens its spread at both, and raises the pass fraction at
\ang{\HstSitASub} from \HstSitAPass{} to \HstSitABrPass{} of \HstSitAN{},
which is the one quantity a mirror seed can buy: on a road that lies on the
far side of the array axis the single branch returns no track while the
detections themselves are correct. The policy therefore follows from which
side of the axis the road lies on, which is known once the siting is chosen.

\textbf{Siting and detection count.} A geometric screen over \num{1112}
candidate positions, applying the mask, range gate, Doppler knee, ambiguity bound
and a two-way loss budget, returns \num{18} feasible sitings, every one limited
by the same constraint: between \SI{41.9}{\percent} and \SI{48.4}{\percent} of
the lap falls below the clutter-filter knee at this speed. None raises the usable
fraction above the \SI{58.1}{\percent} the reported siting already achieves. The
knee is moved by the filter's training length or the dwell; placement does not move it.

\subsection{Coverage of one transmitter--receiver pair}
\label{sec:coverage}
A bistatic pair does not cover a disc around its receiver. Detectability is bounded by
the excess bistatic range $\Delta\rho = R_{tx} + R_{rx} - L$, so the covered region lies
between two confocal ellipses with the transmitter and receiver at the foci: inside by
the \SI{24.4}{\metre} direct-path mask, outside by the \SI{85}{\metre} range gate.
Applying the detector's gates on a \SI{10}{\metre} grid over 36 target headings at
\SI{5.56}{\metre\per\second} gives, for the three receiver positions that synchronise,
covered areas of \SI{4.49}{}, \SI{4.01}{} and \SI{3.32}{\hectare} with target-to-receiver
ranges spanning \SIrange{15}{352}{\metre}. The area is set by the pair's geometry; covering more ground requires more pairs.

This construction states where detection is geometrically admissible; it does not predict how
often the target is detected there. Checked against the five receiver positions whose
outcomes are known, it marks 16 to 18 of the lap's 31 coherent intervals geometrically
usable at every one of them --- both ends of the measured coverage range
(\SIrange{\HstSitHCov}{\HstSitFCov}{\percent}). Geometry excludes positions and does not rank the usable ones; a coverage polygon
drawn from geometry alone overstates what the system delivers.

\subsection{Cost of establishing each reliability figure}
\label{sec:reliability}
Four quantities have to be separated when a coverage claim is made for this class of
system, and they are established at different cost.

\textbf{Geometric admissibility.} This costs seconds and is exact: a position either satisfies
the mask, gate, Doppler knee and ambiguity bound or it does not.

\textbf{Detection probability.} This requires cells. On this scene $P_D$ per coherent interval
is \numrange{0.28}{0.35} with the target-absent null at zero, and it varies by excess-range
band rather than uniformly over the covered area.

\textbf{Tracking coverage.} This requires both nulls of Section~\ref{sec:nullgates} and
\SI{15}{} seeds per position to bound: across the nine sitings of Table~\ref{tab:siting} it
spans \SIrange{\HstSitHCov}{\HstSitFCov}{\percent} with per-position spreads of
\SIrange{\HstSitECovSd}{\HstSitFCovSd}{\percent}, so two positions are separable only
when the difference between them exceeds that spread, which no adjacent pair on the line does.

\textbf{Intervals.} An interval on each requires $n$. Fifteen cells differing only in noise seed
give a standard error of \SI{1.7}{pp} on coverage; five cells gave a standard deviation
of \SI{16}{pp} and could not test the point prediction registered against them. Cells
must also be scored over a matched window: an arm whose cells replayed the target path a
different number of times reported a pass fraction of 4 of 15 where a matched window
gives 2 of 15.

What simulation cannot settle is the absolute level: clutter, co-channel interference,
receiver clock and gain behaviour are the measurement's alone. One field capture with the
mast surveyed and the vehicle on GPS anchors the geometry end to end, by the criterion
already registered --- the detector's reported range must track the GPS-derived excess
bistatic range within one range grid bin (\SI{\BthreeCrsGridDeltaRM}{\metre}) across a lap.

\begin{table}[!t]
\caption{The \SI{\HaLapM}{\metre} High St lap at two target heights beside the
\SI{315}{\metre} Pender St siting re-run with the same taper and seeds, $n = \HaN$
cells differing only in noise seed, $\pm$ one standard deviation, cell scorer at
a \SI{20}{\metre} gate over one pass of the lap. Margin is coverage less the
arm's own permutation null; decoy margin is the range of coverage less the worst decoy; null is
the coverage of the target-absent re-render.}
\label{tab:canyon}
\centering
\footnotesize
\setlength{\tabcolsep}{2.5pt}
\begin{tabular}{lccc}
\toprule
 & High St & High St & Pender St \\
 & vehicle & UAV & reference \\
\midrule
target height [m] & \HaVehZ & \HaUavZ & \LtVehZ \\
lap subtense at RX [deg] & \HaSubDeg & \HaSubDeg & \HaRefSubDeg \\
coverage [\%] & $\HaVehCov \pm \HaVehCovSd$ & $\HaUavCov \pm \HaUavCovSd$ & $\HaRefCov \pm \HaRefCovSd$ \\
permutation null [\%] & $\HaVehPerm \pm \HaVehPermSd$ & $\HaUavPerm \pm \HaUavPermSd$ & $\HaRefPerm \pm \HaRefPermSd$ \\
margin [pp] & $\HaVehMargin \pm \HaVehMarginSd$ & $\HaUavMargin \pm \HaUavMarginSd$ & $\HaRefMargin \pm \HaRefMarginSd$ \\
decoy margin [pp] & \HaVehMarginLo--\HaVehMarginHi & \HaUavMarginLo--\HaUavMarginHi & \HaRefMarginLo--\HaRefMarginHi \\
$P_\mathrm{D}$ & $\HaVehPd \pm \HaVehPdSd$ & $\HaUavPd \pm \HaUavPdSd$ & $\HaRefPd \pm \HaRefPdSd$ \\
IDF1 & $\HaVehIDF \pm \HaVehIDFSd$ & $\HaUavIDF \pm \HaUavIDFSd$ & $\HaRefIDF \pm \HaRefIDFSd$ \\
IDSW & \HaVehIdswLo--\HaVehIdswHi & \HaUavIdswLo--\HaUavIdswHi & \HaRefIdswLo--\HaRefIdswHi \\
RMSE [m] & $\HaVehRmse \pm \HaVehRmseSd$ & $\HaUavRmse \pm \HaUavRmseSd$ & $\HaRefRmse \pm \HaRefRmseSd$ \\
null [\%] & \HaVehNullCov & \HaUavNullCov & \HaRefNullCov \\
\bottomrule
\end{tabular}
\end{table}

\begin{figure*}[t]
\centering
\includegraphics[width=\textwidth]{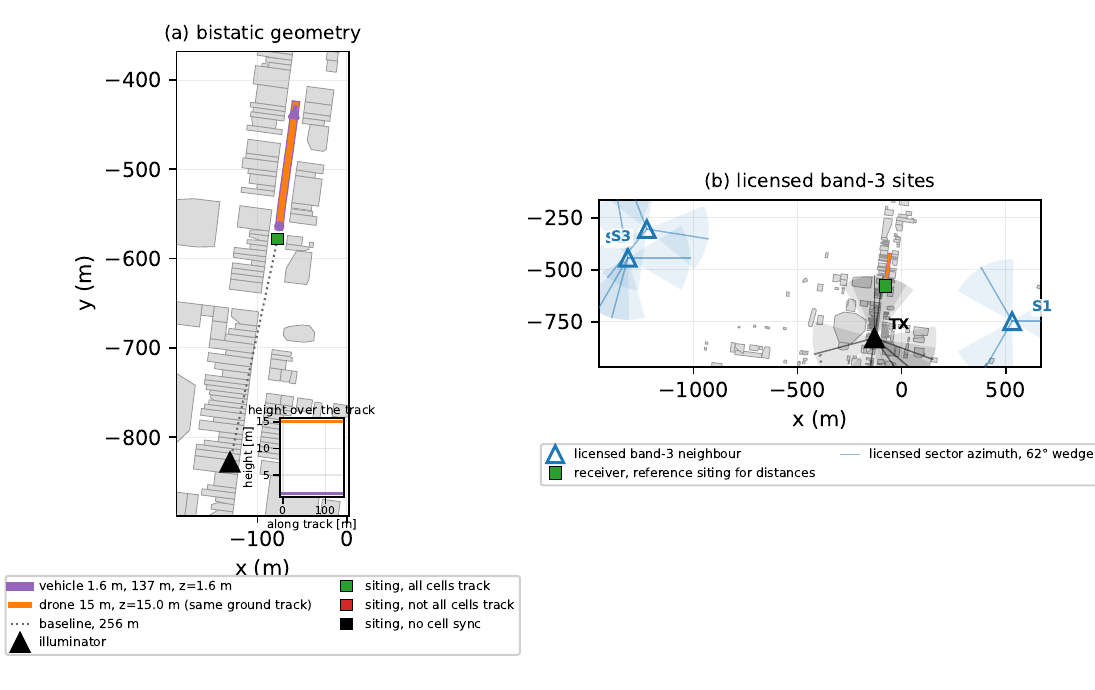}
\caption{The \SI{\HaLapM}{\metre} High St lap with the receiver on the street
centreline, from the scene's meshes, the rendered trace and the ACMA licence
records: (a) the bistatic geometry, with the building footprints that line
both kerbs and the two target heights inset; (b) the illuminator \HaTxSid{} and
the three nearest licensed band-3 sites with their sector azimuths. The lap
subtends \ang{\HaSubDeg} at the receiver.}
\label{fig:canyon_layout}
\end{figure*}

\subsection{A built-up section of High St}
\label{sec:canyon}
Every lap above runs on Pender St, where the nearest building footprint is
\SI{178}{\metre} from the track. This subsection moves the lap onto High St
itself, where the footprints line both kerbs, and asks whether detection and
tracking survive the change from open ground to a street canyon with the
detector, the tracker, the scorer and the operating point held fixed. The
illuminator is a different licensed site from the one that serves the Pender
St laps (site S1 of Table~\ref{tab:sites}, which is the nearer of the two to
this section of High St): site \HaTxSid{} at
$(\HaTxX, \HaTxY, \HaTxZ)$~\si{\metre}, \SI{\HaTxOffCentreM}{\metre} west of the
street centreline, rendered on the sector that faces along the street; the
receiver is a stationary UE at $(\HaUeX, \HaUeY, \HaUeZ)$~\si{\metre} on the
same centreline, \SI{\HaBaseline}{\metre} from the illuminator, with its
array boresight across the street. The lap is \SI{\HaLapM}{\metre} of the
centreline, which bears \ang{\HaStreetBearing}, driven one way at no more than
\SI{\HaSpeedMax}{\metre\per\second}; the vehicle arm places the target at
\SI{\HaVehZ}{\metre} and the airborne arm at \SI{\HaUavZ}{\metre} over the same
ground track. The carrier is the site's licensed \SI{\HaFcMHz}{\mega\hertz}
rather than the \SI{\HstFcMHz}{\mega\hertz} of the Pender St laps; on the Pender
geometry that change moves coverage by \SI{\HaCarrierDeltaPp}{pp}
($t = \HaCarrierDeltaT$), which is not separable from seed at $n = \HaN$ and is
declared and left uncorrected. The coherent interval, range gate, Doppler gate
and range taper are those of Table~\ref{tab:longtraj}.
Figure~\ref{fig:canyon_layout} draws the geometry.

Table~\ref{tab:canyon} reports both arms over $n = \HaN$ cells differing only
in noise seed, beside a reference arm that is the \SI{315}{\metre} Pender St siting
of Section~\ref{sec:siting} re-run on the same seeds with the occupied-band
taper. The reference is re-run because the siting figures of
Table~\ref{tab:siting} were measured with the full-grid taper, under which the
same siting reads \SI{\HaRefCovTaZero}{\percent} $\pm$
\SI{\HaRefCovTaZeroSd}{\percent}; pairing the canyon against that number would
charge about \SI{40}{pp} of taper to the geometry. All three arms are scored
over one pass of the lap, cut at the first replay of the trajectory, because
the driver's own scoring window admits the replay and credits it.

Tracking is demonstrated on both canyon arms on every cell: coverage exceeds
its own permutation null by \SI{\HaVehMargin}{pp} $\pm$ \SI{\HaVehMarginSd}{pp}
for the vehicle and \SI{\HaUavMargin}{pp} $\pm$ \SI{\HaUavMarginSd}{pp} for the
airborne target, and exceeds the worst decoy by
\SIrange{\HaVehMarginLo}{\HaVehMarginHi}{pp} on the vehicle cells and
\SIrange{\HaUavMarginLo}{\HaUavMarginHi}{pp} on the airborne cells.

The permutation null and the identity switches share one geometric cause,
which the layout measures directly. The lap subtends \ang{\HaSubDeg} of azimuth
at the receiver against \ang{\HaRefSubDeg} at the reference siting, because the
receiver stands on the street axis and the target runs almost directly away
from it; the bistatic range rate is \HaRhoDotOverV{} times the ground speed on
this trace, so the measurement is close to range-only. A $1\times 2$ array has
no azimuth leverage on a radial track, so any track placed along that radial
covers the truth under the scorer's gate, which is what a permutation null of
\SIrange{61}{67}{\percent} records, and the tracker cannot hold an identity
across the detection gaps that a range-only measurement leaves.

That attribution was tested by moving the receiver \SI{\HoOffsetM}{\metre}
east of the centreline, onto the kerb line, which raises the subtense to
\ang{\HoSubDeg} for \SI{\HoEchoDb}{\decibel} of echo and leaves the line of
sight to every snapshot of the lap. Over $n = \HoN$ cells paired by seed with
the centreline receiver, the identity switches halve, \HoIdsw{} $\pm$
\HoIdswSd{} against \HoRefIdsw{} $\pm$ \HoRefIdswSd{} per cell, lower on
\HoN{} of \HoN{} pairs; the position error rises by \SI{\HoDRmse}{\metre}
on \HoDRmsePos{} of \HoN{} pairs; and the identity score, \HoIDF{} $\pm$
\HoIDFSd{}, the detection rate, \HoPd{} $\pm$ \HoPdSd{}, the coverage,
\SI{\HoCov}{\percent} $\pm$ \SI{\HoCovSd}{\percent}, and the margin over
the permutation null, \SI{\HoMargin}{pp} $\pm$ \SI{\HoMarginSd}{pp}, do not
separate from the centreline values at this $n$ (the coverage spread carries
one cell at \SI{\HoOutlierCov}{\percent}). On this lap subtense improves identity
continuity and leaves identity score and detections within their spread; the chain of
Section~\ref{sec:chain} is built on the centreline.

\begin{figure*}[!t]
\centering
\includegraphics[width=0.8\textwidth]{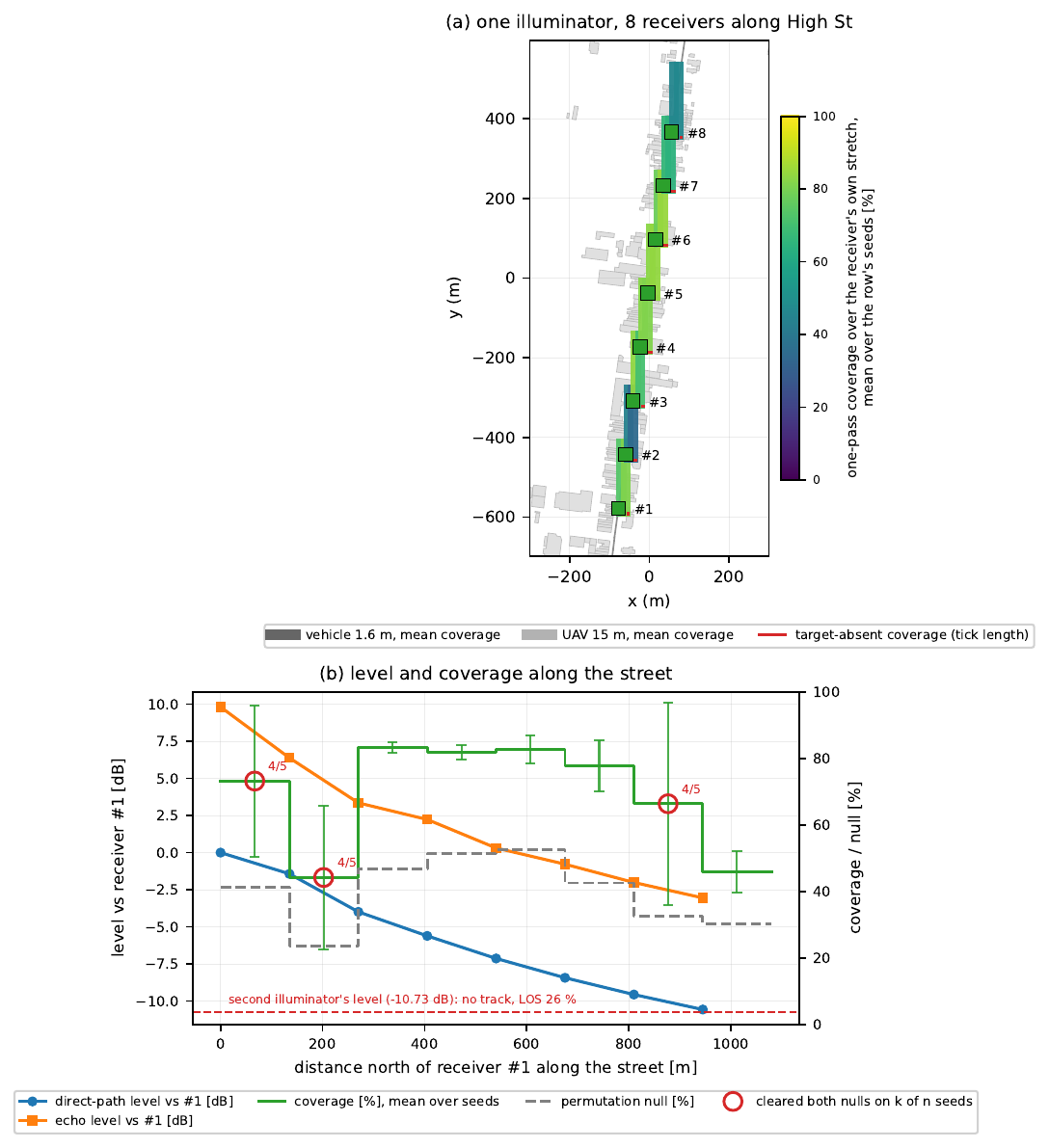}
\caption{One illuminator and \HcxN{} receivers along High St: (a) the chain
over the building footprints, each receiver's stretch drawn as two strips
coloured by the mean one-pass coverage over seeds for the vehicle (left) and
the airborne target (right), with the target-absent coverage as a tick whose
length is its value; (b) direct-path and echo level relative to the first
receiver, mean vehicle coverage with its standard deviation over seeds and the
permutation null per stretch against distance north of the first receiver,
rows that cleared both nulls on $k$ of $n$ seeds ringed, and the direct-path
level of the failed second illuminator of Section~\ref{sec:phasec} marked.}
\label{fig:chain}
\end{figure*}

\begin{table*}[!t]
\caption{The \SI{\HcxSpanM}{\metre} chain, mean $\pm$ standard deviation over
\HcxSeedsLo{} noise seeds per receiver and target, cell scorer at a
\SI{20}{\metre} gate over one pass of each receiver's lap, with
$\Delta P_\mathrm{dir}$ the rendered direct-path level relative to receiver 1,
margin the coverage less the permutation null, and absent the coverage of the
target-absent cells. A superscript $k/n$ marks a row that exceeded both of its
nulls on $k$ of $n$ seeds, so its mean and standard deviation span two modes;
the absent column carries none because $k=0$ is its required outcome.}
\label{tab:chain}
\centering
\footnotesize
\begin{tabular}{@{}rrr rrr rrr r@{}}
\toprule
\# & $y$ & $\Delta P_\mathrm{dir}$ & \multicolumn{3}{c}{vehicle, \SI{1.6}{\metre}} & \multicolumn{3}{c}{UAV, \SI{15}{\metre}} & absent \\
\cmidrule(lr){4-6}\cmidrule(lr){7-9}
 & [m] & [dB] & cov. & margin & $P_\mathrm{D}$ & cov. & margin & $P_\mathrm{D}$ & cov. \\
 & & & [\%] & [pp] & & [\%] & [pp] & & [\%] \\
\midrule
1 & -578 & +0.0 & 73.2\,$\pm$\,22.7\textsuperscript{4/5} & +32.0 & 0.21 & 81.1\,$\pm$\,5.6 & +38.2 & 0.26 & 4.1\,$\pm$\,1.9 \\
2 & -443 & -1.4 & 44.2\,$\pm$\,21.6\textsuperscript{4/5} & +12.7 & 0.22 & 34.7\,$\pm$\,6.8\textsuperscript{1/5} & +4.6 & 0.16 & 10.2\,$\pm$\,3.2 \\
3 & -308 & -4.0 & 83.2\,$\pm$\,1.6 & +35.4 & 0.33 & 70.5\,$\pm$\,26.9\textsuperscript{4/5} & +25.5 & 0.34 & 11.1\,$\pm$\,11.9 \\
4 & -173 & -5.6 & 81.8\,$\pm$\,2.1 & +42.3 & 0.27 & 81.2\,$\pm$\,3.5 & +44.6 & 0.28 & 17.1\,$\pm$\,7.3 \\
5 & -38 & -7.1 & 82.7\,$\pm$\,4.2 & +34.0 & 0.30 & 83.9\,$\pm$\,1.9 & +31.8 & 0.30 & 0.0\,$\pm$\,0.0 \\
6 & 97 & -8.4 & 77.7\,$\pm$\,7.7 & +39.2 & 0.27 & 84.7\,$\pm$\,1.2 & +43.8 & 0.28 & 20.9\,$\pm$\,12.7 \\
7 & 232 & -9.6 & 66.4\,$\pm$\,30.4\textsuperscript{4/5} & +32.8 & 0.22 & 64.1\,$\pm$\,33.4\textsuperscript{4/5} & +31.0 & 0.23 & 14.2\,$\pm$\,0.9 \\
8 & 367 & -10.6 & 45.9\,$\pm$\,6.2 & +15.7 & 0.17 & 46.7\,$\pm$\,21.9\textsuperscript{3/5} & +12.3 & 0.24 & 2.0\,$\pm$\,3.5 \\
\bottomrule
\end{tabular}

\end{table*}

\section{Coverage from independent receiver segments along one kilometre of street}
\label{sec:chain}
One receiver on the street axis covers a stretch bounded on the near side by
the direct-path mask and on the far side by the range gate; on this geometry
that stretch is \SI{\HcxStretchM}{\metre}, and it lies north of the receiver
only, because the street runs nearly along the baseline and the southern side
is the bistatic blind zone. Covering a longer length of street with one
illuminator is therefore a question of how many receivers, at what spacing,
and whether the pairs further from the illuminator still track. This section
places \HcxN{} receivers along High St at \SI{\HcxSpacingM}{\metre} spacing,
from the receiver of Section~\ref{sec:canyon} at $y = \HcxYfirst$~m to $y =
\HcxYlast$~m, and runs the vehicle and the airborne target past each. The
\SI{\HcxOverlapM}{\metre} overlap between adjacent stretches is chosen to
exceed the distance the target covers while the tracker's birth rule
accumulates its first confirmations. The chain spans \SI{\HcxSpanM}{\metre}
of street from the first stretch's start to the last stretch's end.

Each receiver is run as its own cell, with the target driven over that
receiver's stretch plus \SI{\HcxOverlapM}{\metre} either side
(\SI{\HcxLapLenM}{\metre} at \SI{\HaSpeedMax}{\metre\per\second}) on waypoints
taken from one global parametrisation of the whole street, so every cell's
record carries global time and position and the chain is assembled offline by
time ordering alone. No detection or track is fused across receivers and no
identity is carried over a handover; each receiver's cell is replayed in
isolation, so the chain is a set of independent local trials assembled on a
common clock; simultaneous reception and continuous
tracking across receivers are not demonstrated, and mutual interference between receivers, which
a passive receiver does not create, and the shared-illuminator timing that
simultaneous reception would require are not represented. Every receiver is screened on its
rendered trace before a cell is run: all \HcxN{} hold line of sight to the
target on every snapshot, the direct path falls to \SI{\HcxDirectLoDb}{\decibel}
relative to the first receiver at the eighth, and the echo-to-direct ratio is
\SIrange{\HcxRatioLoDb}{\HcxRatioHiDb}{\decibel} better than the first
receiver's at every other one, because each lap begins
\SI{\HcxOverlapM}{\metre} before its own receiver rather than
\SI{150}{\metre} from it. Three target-absent cells are run per receiver.

Figure~\ref{fig:chain} and Table~\ref{tab:chain} report \HcxSeedsLo{} noise
seeds per receiver and target, \HcxVehCells{} vehicle cells and
\HcxUavCells{} airborne cells over the \HcxGood{} receivers. The vehicle's
coverage exceeds both of its nulls on \HcxVehK{} of those \HcxVehCells{} cells
and on every seed at \HcxVehDem{} of the \HcxGood{} receivers, with per-receiver
means over the receiver's own stretch between \SI{\HcxVehCovLo}{\percent} and
\SI{\HcxVehCovHi}{\percent}, a length-weighted mean of
\SI{\HcxVehCovW}{\percent}, and a per-receiver standard deviation over seeds
between \SI{\HcxVehSdLo}{\percent} and \SI{\HcxVehSdHi}{\percent}. The airborne
target's does so on \HcxUavK{} of \HcxUavCells{} cells and on every seed at
\HcxUavDem{} receivers, with a length-weighted mean of
\SI{\HcxUavCovW}{\percent}. Receivers \HcxBimodalRx{} are the rows where the
outcome is not the same on every seed: they pass on some seeds and fail the
decoy or permutation null on others, so the standard deviation there spans two
modes and is not a spread, and the coverage of the seeds that passed
is \SIrange{\HcxUavCovPassLo}{\HcxUavCovPassHi}{\percent} for the airborne
target. Detection rate per receiver is
\numrange{\HcxVehPdLo}{\HcxVehPdHi} for the vehicle and
\numrange{\HcxUavPdLo}{\HcxUavPdHi} for the airborne target, at the registered
$\pm$1 range cell and $\pm$1 Doppler bin gate. The eighth receiver, whose direct path is at the
level at which the second illuminator of Section~\ref{sec:phasec} failed,
clears both nulls for both targets with full line of sight; that failure is
therefore not attributable to the direct-path level alone. The target-absent
cells cover between \SI{\HcxNullLo}{\percent} and \SI{\HcxNullHi}{\percent} of
the truth with no detection on the target, \HcxNullAboveFive{} of \HcxGood{}
above \SI{5}{\percent}, which is the gate-occupancy null of
Section~\ref{sec:nullgates} at a \SI{300}{\metre} gate, and each coverage
figure in the table is read against its own row's value. None of the absent
cells demonstrates tracking on any seed.

The seed-to-seed spread is not uniform along the chain, and it separates two
regimes. Where the vehicle clears both nulls on every seed, the standard
deviation over seeds is
\SIrange{\HcxVehSdAllPassLo}{\HcxVehSdAllPassHi}{\percent} of coverage, so
the per-receiver figure is reproducible at that receiver. At receivers
\HcxVehBimodalRx{} it does not clear them on every seed, and the standard
deviation reaches \SI{\HcxVehSdHi}{\percent} because the row holds a mode that
tracks and a mode that does not; the airborne rows behave the same way at
receivers \HcxBimodalRx{}. The figure to read for those rows is the coverage of
the seeds that cleared both nulls,
\SIrange{\HcxVehCovPassLo}{\HcxVehCovPassHi}{\percent} for the vehicle, with the
pass fraction beside it in the table. The chain's claim is the length of street
over which the vehicle clears both nulls on every seed --- \HcxVehDem{} of
\HcxGood{} receivers --- and the value in any single row is secondary to it. The intervals
of Table~\ref{tab:canyon} do not transfer to these rows: the chain states
where one illuminator's coverage ends along the street; the reliability at any
one receiver is given by that receiver's own row.

\subsection{Field bounds on the chain}
\label{sec:chainfield}
The simulated receiver is not noise-limited (Section~\ref{sec:limits}), so
the question a deployment asks --- at which receiver along the street the echo
falls into the noise --- is answered from a link budget, since the cells do not reach the noise floor. Table~\ref{tab:linkbudget} gives it per receiver: free-space direct
path and echo from the licensed EIRP of \SI{\LbEirpDbm}{dBm}, a point target
of \SI{\LbRcsDbsm}{dB} re \SI{1}{\square\metre}, unity receive gain, and a
thermal floor of \SI{\LbThermalDbm}{dBm} in \SI{\LimThermalBwMHz}{\mega\hertz}
at a \SI{\LbNfDb}{\decibel} noise figure, with
\SI{\LbCohGainDb}{\decibel} of coherent gain, $10\log_{10}(\LbReN \times
\LbOccN)$, from \LbReN{} reference resource elements per occasion coherently
combined over \LbOccN{} occasions, an upper bound that assumes a stationary
echo over the interval and no loss from the taper. At the eighth receiver,
\SI{\LbRxEightBaselineM}{\metre} from the illuminator, the echo at the far
end of the lap is \SI{\LbRxEightEchoEndDbm}{dBm}, \SI{\LbRxEightSnrEndDb}{\decibel}
above the floor after integration and \SI{\LbRxEightMarginDb}{\decibel} above
the detector's threshold trim; the direct path there is
\SI{\LbRxEightDirDbm}{dBm}, \SI{\LbRxEightDpiOverEchoDb}{\decibel} above
the echo. Under these assumptions no receiver of the chain is bounded by the
thermal floor, and the quantity that binds is the direct-path-to-echo ratio,
which is the limit the simulated arm already operates at. The budget is free
space with a point target and no multipath, building or clutter loss, and is
a bound: a deployment will sit below it.

Two further measurements bound how the chain can be joined. The
\SI{300}{\metre} range gate ends each receiver's detections
\SI{\GfMissM}{\metre} short of where the next receiver's begin, so no truth
position is seen by two receivers at once and no measurement can be fused
across a handover at this spacing. Widening the gate to \SI{\GfGateM}{\metre}
at receiver \GfRx{}, with every other setting and the same five seeds held,
opens a \SI{\GfJointBandM}{\metre} band seen by both receivers, raises the
detection rate from \num{\GfPdThree} $\pm$ \num{\GfPdThreeSd} to
\num{\GfPdFour} $\pm$ \num{\GfPdFourSd}, and lowers coverage from
\SI{\GfCovThree}{\percent} $\pm$ \SI{\GfCovThreeSd}{\percent}, both nulls
cleared on \GfKThree{} of 5 seeds, to \SI{\GfCovFour}{\percent} $\pm$
\SI{\GfCovFourSd}{\percent} on \GfKFour{} of 5, with the reported-object
count rising from \GfRepThree{} to \GfRepFour{} per cell; the chain keeps the
\SI{300}{\metre} gate. In the opened band, \GfCoinc{} truth frames over the
five seeds carry a detection from both receivers; locating the target on the
street line (the line is the third constraint; two bistatic ranges alone fix
a curve rather than a point) from one receiver's bistatic range gives an error of
\SI{\GfOneRxM}{\metre} $\pm$ \SI{\GfOneRxSdM}{\metre}, from the two
receivers' ranges combined \SI{\GfTwoRxM}{\metre} $\pm$
\SI{\GfTwoRxSdM}{\metre}, against a permutation null of
\SIrange{\GfPermLoM}{\GfPermHiM}{\metre}: both beat the null and the two are
indistinguishable. Half a range grid bin projected onto the street is
\SI{\GfRangeCellProjM}{\metre}, of the same order as the single-receiver
error.

\begin{table}[t]
\caption{Field link budget per chain receiver, free space from the licensed
EIRP with a \SI{\LbRcsDbsm}{dB} re \SI{1}{\square\metre} point target and
unity receive gain; SNR is after \SI{\LbCohGainDb}{\decibel} of coherent gain
against a \SI{\LbThermalDbm}{dBm} floor. The budget is an upper bound: no
multipath, building or clutter loss is included.}
\label{tab:linkbudget}
\centering
\footnotesize
\begin{tabular}{@{}rrr rr rr@{}}
\toprule
\# & baseline & $P_\mathrm{dir}$ & \multicolumn{2}{c}{echo, lap start} & \multicolumn{2}{c}{echo, lap end} \\
\cmidrule(lr){4-5}\cmidrule(lr){6-7}
 & [m] & [dBm] & [dBm] & SNR [dB] & [dBm] & SNR [dB] \\
\midrule
1 & 256 & -20.1 & -46.9 & +86.6 & -70.6 & +62.9 \\
2 & 392 & -23.8 & -50.8 & +82.7 & -73.0 & +60.5 \\
3 & 528 & -26.4 & -53.5 & +79.9 & -74.8 & +58.7 \\
4 & 664 & -28.4 & -55.6 & +77.9 & -76.4 & +57.1 \\
5 & 800 & -30.0 & -57.3 & +76.2 & -77.7 & +55.8 \\
6 & 937 & -31.4 & -58.7 & +74.8 & -78.8 & +54.7 \\
7 & 1073 & -32.5 & -59.9 & +73.6 & -79.8 & +53.7 \\
8 & 1209 & -33.6 & -61.0 & +72.5 & -80.7 & +52.8 \\
\bottomrule
\end{tabular}

\end{table}

\section{Multiple illuminators at one receiver}
\label{sec:multipci}
A single transmitter--receiver pair measures excess bistatic range, range rate
and a cone angle about the receiver array axis; it does not measure elevation,
and the tracks of Section~\ref{sec:longtraj} are therefore confined to the
intersection of the target-height plane with the constant-bistatic-range
surface, with the height taken from the scene; the planar description is an
approximation that holds while the transmitter and receiver heights differ
from the target's by less than the baseline. In the uplink testbed the elevation was
recovered by a second transmitter~\cite{gurung-srs-oran-2026}. On the downlink
the same role is taken by the co-channel cells licensed around the receiver:
each cell identity received at one user-equipment position forms its own
bistatic pair, self-referenced to that cell's own direct path; two
constant-bistatic-range surfaces intersect in a curve, and a third pair, or
the target-height plane used as a prior in this paper, is needed to fix a
point, so the dilution computed below is that of three or more surfaces or of
two surfaces with the height plane, as stated per row. This
section states the configuration under construction and the geometric dilution
it affords.

\subsection{One cell identity per measurement}
The emulated receiver holds synchronisation on one cell identity. With
\MpResyncIds{} identities present in the emulated channel the receiver
re-synchronised \MpResyncN{} times over the run and produced no usable
detection stream, so the multi-illuminator measurement is built from
single-identity cells rendered on one trace timeline and fused offline, one
illuminator per cell. In the field the restriction does not apply: cells on the
same carrier arrive in the same samples, and one capture at one receiver yields
one pair per co-channel identity, each with its own frame timing and delay
origin, with five of every six identities placing their reference signal on a
different subcarrier comb from the serving cell's (Section~\ref{sec:highst}).
The field route is therefore per-identity extraction on a single recording,
and the simulated route is its one-identity-per-cell equivalent.

\subsection{Illuminators available at the site}
Table~\ref{tab:sites} decides which licensed sites can contribute. A neighbour
is a usable pair only if it is co-channel with the serving cell and one of its
sectors faces the receiver; at this site the nearest neighbour is
\SI{51}{\degree} off its best sector, the strongest carries no co-channel
component, and the only neighbour that is both co-channel and within
\SI{30}{\degree} of boresight is the furthest, at \SI{1177}{\metre} with a
free-space direct-path bound of \SI{-38}{dBm}. The serving site itself is
three-sectored; the sector reported in Section~\ref{sec:highst} is
\SI{19}{\degree} off the receiver and the other two present their back lobes
to it. Whether a cell search on the live band-3 carrier returns the furthest
neighbour at a kerbside receiver is not answerable in simulation and is deferred
to the first recording.

\subsection{Geometric dilution of the candidate configurations}
Table~\ref{tab:multipci} gives the vertical dilution of precision of three
receiver configurations against the \MpCoChannelSectors{} co-channel sectors
over \MpSites{} licensed sites of the scene, from a bistatic-range-only Jacobian
with equal measurement error per pair, an azimuth wedge of \ang{32.5} either side
of each sector's boresight and no link budget; the medians are taken over the
grid cells that every configuration can solve. One receiver with the licensed
transmitters resolves altitude at \SI{\MpAirZ}{\metre} with a dilution of
\MpVdopAirA{}; a second receiver at a different height lowers it to
\MpVdopAirD{} and more than doubles the solvable fraction of the grid, and a
third receiver adds coverage rather than lowering dilution. Forcing every transmitter to
one mast height leaves the one-receiver dilution at \MpVdopAirAflat{}, so the
vertical sensitivity comes from the transmitters' spread in azimuth and range,
not from their heights. Three receivers at a common height is the worst
configuration computed, because the three pairs share both the transmitter leg
and the receiver height. The dilution treats each transmitter--receiver pair
as an independent range with equal error; co-sited sectors of one licensed
site share a mast and a clock, so their reference errors are correlated and
the dilution understates the vertical error where such pairs dominate.

\begin{table}[t]
\caption{Vertical dilution of precision (VDOP, median over the commonly solvable
grid) and solvable fraction of the \SI{1900}{\metre} tile for three receiver
configurations against the co-channel sectors of Table~\ref{tab:sites}, from
geometry alone, computed with no link budget and equal error per pair.}
\label{tab:multipci}
\centering
\footnotesize
\setlength{\tabcolsep}{4pt}
\begin{tabular}{@{}lccc@{}}
\toprule
receivers & solvable & VDOP & VDOP \\
($N$ transmitters) & [\%] & at \SI{1.6}{\metre} & at \SI{\MpAirZ}{\metre} \\
\midrule
1 & \MpCoverA & \MpVdopGroundA & \MpVdopAirA \\
2, staggered height & \MpCoverD & \MpVdopGroundD & \MpVdopAirD \\
3, staggered height & \MpCoverC & \MpVdopGroundC & \MpVdopAirC \\
\bottomrule
\end{tabular}
\end{table}

\subsection{Measurement plan and gates}
The first configuration is one receiver with the serving sector and each usable
co-channel sector in turn, on the \SI{\LtLapM}{\metre} lap of
Section~\ref{sec:longtraj}; each illuminator carries its own alignment record
(boresight, element gain and excess-range support) and a sector that fails that
record is reported and not run. Fusion is at the detection level with a
three-dimensional state. Two controls gate any fused figure: the fusion of the
target-absent streams of every illuminator must produce no track, and the fusion
of repeated copies of one illuminator's stream must reproduce that
illuminator's single-pair result. The quantities to be reported are the
single-pair and fused identity, switch and GOSPA figures of
Section~\ref{sec:metrics} on the same cells, and the altitude error of the fused
track against the rendered height, which no single pair here can report.

\subsection{A second illuminator at the street receiver}
\label{sec:phasec}
One pair of that plan has been run, at the High St receiver of
Section~\ref{sec:canyon}: licensed site \HcSid{} at \SI{\HcFcMHz}{\mega\hertz},
\SI{\HcDistM}{\metre} from the receiver and \ang{\HcOffBoreDeg} off its
nearest sector, as a separate cell because its carrier is \SI{30}{\mega\hertz}
from the primary's. Its direct path arrives \SI{\HcDirectRelDb}{\decibel}
relative to the primary's and its echo-to-direct ratio over the lap is within
\SI{\HcRatioRelDb}{\decibel} of the primary's, so the pair passes a ratio
screen. The rendered trace carries the echo in \HcLosSnap{} of \HcSnapN{}
snapshots (\SI{\HcLosPct}{\percent}), the northern third of the lap being
shadowed by the frontage from this transmitter's direction, and the cell
detects the target in \HcDet{} of \HcObs{} opportunities ($P_\mathrm{D} =
\HcPd$), with coverage \SI{\HcCov}{\percent}, a permutation null of
\SI{\HcPerm}{\percent} and a worst decoy of \SI{\HcWorstDecoy}{\percent}:
under the confirmed-track convention the arm scores nothing at all, so no
track is claimed and no null comparison is made for it. At the same trim it
publishes \HcNdet{} detection records over the run, read from the cell's own
stream, against the primary illuminator's \HcNdetPrimary{} $\pm$
\HcNdetPrimarySd{} over \HcNdetPrimaryN{} cells: the occluded illuminator
emits more records while detecting the target in \HcDet{} of \HcObs{}
opportunities. Three things differ from the primary at once --- the
\SI{\HcLosPct}{\percent} line-of-sight fraction, the lower direct path and
hence a different cancellation residual at the same trim, and the
off-boresight illumination --- and one cell does not separate them. What the
cell establishes is that a ratio screen is not a sufficient admission gate for
a second illuminator, and that at this receiver the pair contributes no
measurable quantity to fuse; the plan above starts from the
line-of-sight fraction of each candidate pair, read from the trace before
any cell is run.

\section{Limitations}
\label{sec:limits}
\textbf{Association gate.} Every coverage figure is a decision at the
\SI{\LimGateM}{\metre} scorer gate, which on this scene thresholds a
continuous track-position error: at a \SI{\LimGateWideM}{\metre} gate the two
coverage populations of \SI{\LimModeHi}{\percent} and \SI{\LimModeLo}{\percent}
converge to \SI{\LimConvLo}{\percent}--\SI{\LimConvHi}{\percent}, and with the
position error removed the same runs reach \SI{\LimCeil}{\percent}. A coverage
figure from this scene is a thresholded position error at a stated gate.

\textbf{Channel sampling within the dwell.} The channel trace is sampled once
per coherent interval, so range migration and channel variation inside a dwell
are not represented and their effect on coverage is not quantified.

\textbf{Illuminator pattern, ports, tilt and polarisation.} The transmitter is
one TR~38.901 element (\ang{\LimElemBwAzDeg} by \ang{\LimElemBwElDeg}) on one
reference-signal port with no vertical array factor and no downtilt, and the
channel model injects a single transmit column, so a deployed panel's
\ang{\LimPanelBwLoDeg}--\ang{\LimPanelBwHiDeg} elevation pattern, its port
count and its slant polarisation are recorded with each configuration and not
applied. The Pender St figures are for the sector \ang{19} off the receiver;
the facing sector lifts coverage by \SI[retain-explicit-plus]{\LimFacingLiftPp}{pp}
(standard error \SI{\LimFacingLiftSePp}{pp}), which is not separable, so the
figures are those of a stated configuration.

\textbf{Elevation geometry of an aerial target.} Over the published panel's
electrical tilt range the echo-to-direct ratio at the receiver changes by
\SI{\PanelRatioFifteenLoDb}{\decibel}--\SI[retain-explicit-plus]{\PanelRatioFifteenHiDb}{\decibel}
for a target at \SI{15}{\metre} and by
\SI{\PanelRatioSixtyLoDb}{\decibel}--\SI{\PanelRatioSixtyHiDb}{\decibel} at
\SI{\LimAerialZ}{\metre}; the \SI{15}{\metre} results transfer to a deployed
panel within about \SI{2}{\decibel} of illumination and the
\SI{\LimAerialZ}{\metre} results do not. The first sidelobe is not modelled and
the quantity bounded is illumination.

\textbf{Channel realism.} The channel carries \SimTwoTapCount{} static taps
(rms delay spread \SI{\SimTwoTapRmsM}{\metre}), a cap of the channel
forwarder, so every statement about the built-up street describes a two-path
rendering of its geometry; clutter, multipath richness and ground-bounce
structure are not represented, and results that depend on the number or
spread of static returns (direct-path mask, CFAR training) do not carry to a
measured channel. A delay-spread estimate from a public live-downlink
recording~\cite{estevez-lte-b20} did not pass its controls and is not quoted.

\textbf{Noise and absolute level.} The receiver is not noise-limited: an
injected tone reads back within \SI{\MthirteenToneErrDb}{\decibel}, but the
receiver's own noise readback (\SI{\MthirteenReportedNzeroDbm}{dBm}, produced
at \texttt{\MthirteenNzeroOrigin}) is \num{\MthirteenDigitalNoiseLsb}~LSB of
amplitude, \SI{\MthirteenFloorGapDb}{\decibel} from $kT_0B$ at the declared
\SI{\MthirteenNfDb}{\decibel} noise figure over
\SI{\MthirteenBwMHz}{\mega\hertz}, and the configuration readback gives
\SI{\LimNoiseRmsLsb}{LSB} rms; the two readbacks disagree and both place the
floor at or below the quantiser. The direct path is carried at a normalised
level, \SI{\LimDirectHeadroomDb}{\decibel} below full scale of the
\num{\LimAdcBits}-bit word, whereas a receiver at the free-space bound of
Table~\ref{tab:sites} would hold it at most
\SI{\LimFieldDirectOverNoiseDb}{\decibel} above thermal noise. No range limit
is noise-bounded, no noise-sensitivity claim is made, and no absolute-level
conclusion transfers to a deployment.

\textbf{Doppler margin and dwell.} The unambiguous bistatic range-rate span at
this burst stride is $\pm\SI{8.13}{\metre\per\second}$ against a trajectory
maximum of \SI{3.68}{\metre\per\second}; a faster or more radial target aliases.
Processing every slot removes the bound and shortens the dwell to
\SI{64}{\milli\second}, which requires a trace re-rendered at that snapshot
period and a threshold trim derived at that dwell; neither is reported here.

\textbf{Scene, receiver position, height and speed.} One target on one
straight trajectory per scene; nothing bears on multiple targets. A receiver
at half the baseline, two positions outside the directional illumination and
heights above \SI{7}{\metre} (where the user equipment does not synchronise)
are not measurable with this arm, and the half-baseline comparison rests on
detection alone (\BthreeHalfLMdp\ against \BthreeGateMidMdp). Detection
depends on target speed (missed-detection probability \SpeedThreeMdp,
\SpeedSevenMdp\ and \BthreeCrsBankedMdp\ at \SI{3}{}, \SI{7}{} and
\SI{10}{\kilo\metre\per\hour}; no detection at \SI{25}{\kilo\metre\per\hour},
where part of the range rate falls outside the unambiguous span); no quantity
is corrected for it, and both arms share the trajectory.

\textbf{Scan volume.} Widening the range gate from \BthreeCrsMaxR~m to
\BthreeGateMidMaxR~m and \BthreeGateWideMaxR~m raises the type-2 false-alarm
probability from \BthreeCrsBankedFap\ to \BthreeGateMidFap\ and
\BthreeGateWideFap\ (single cells), so receiver-position figures are quoted
against the \BthreeGateMidMaxR~m control.

\textbf{Scoring convention.} Every coverage figure is scored on confirmed
tracks with one scorer version. Under the earlier convention, which counted
every published record, the second receiver implementation
(Section~\ref{sec:altrx}), the half-baseline position, the two range-gate
controls and two earlier rounds of arm~A read
\SI{\DiffLoCovVone}{\percent}--\SI{\DiffHiCovVone}{\percent} with every cell
passing; under the present one they read
\SI{\DiffLoCovVtwo}{\percent}--\SI{\DiffHiCovVtwo}{\percent} with none passing,
and the two headline arms of Table~\ref{tab:track} moved by
\SI{\DiffCrsShiftPp}{pp} and \SI{\DiffSrsShiftPp}{pp} and kept every cell.
Those families are reported as negative results.

\textbf{Resolution at this $n$.} An arm-to-arm difference below the execution
interval of Section~\ref{sec:variance} (\BthreeSrsExecResolvePd{} of $P_D$,
\SI{\BthreeSrsExecResolveCovPp}{pp} of coverage) is not resolvable at
$n = \BthreeCrsCovN$; the coverage ordering between the arms changes sign
between two rounds on one build (\SI{\BthreeArmDCovDtwo}{pp} and
\SI{\BthreeArmDCovRep}{pp}) and is not claimed.

\textbf{Metric scope and false-alarm bound.} The missed-detection probability
is computed from the same detection flag as the detection probability, so
their identity holds by construction; the coverage and identity quantities
have no 3GPP equivalent. The false-alarm rates are upper bounds from eleven
target-absent replays per arm (about $10^{6}$ scanned cells each) that treat
neighbouring cells as independent trials, and cannot separate two arms that
both meet the declared rate.

\section{Conclusion}
A downlink cell-specific reference signal arm with the receiver at the user
equipment detects and tracks a single ground vehicle and a single airborne
target on the scenes reported, under one receiver implementation, one scorer,
one association gate and one scoring convention held across the paper, with
every coverage figure read against a permutation null, a decoy null and a
target-absent null, and with the configurations that do not demonstrate
tracking under that convention reported beside those that do. On the synthetic canyon its missed-detection probability,
$\BthreeCrsMdp \pm \BthreeCrsMdpSd$ over $n = \BthreeCrsCovN$ noise
realisations, is below that of an uplink sounding reference signal arm on
the same band and target, $\BthreeSrsMdp \pm \BthreeSrsMdpSd$, by
\BthreeArmDPdDtwo{} of detection probability paired by seed; both arms exceed the 3GPP reference
target of 0.05 by an order of magnitude. On the surveyed site, nine receiver
sitings along the street all track the vehicle, at coverage between
\SI{\HstSitHCov}{\percent} and \SI{\HstSitFCov}{\percent}; target
height between \SI{\LtVehZ}{} and \SI{\LtUavBZ}{\metre} changes identity more
than detection; and the built-up street itself costs detection rate and
identity against open ground while leaving coverage unchanged; a receiver
displaced off the street axis halves the identity switches and changes
nothing else. Eight receivers at \SI{\HcxSpacingM}{\metre} along the street
carry one illuminator's coverage over \SI{\HcxSpanM}{\metre}: the vehicle
clears both nulls on \HcxVehK{} of \HcxVehCells{} cells and on every seed at
\HcxVehDem{} of the \HcxN{} receivers, the airborne target on \HcxUavK{} of
\HcxUavCells{}, and a free-space link budget places every receiver of the
chain above the thermal floor, so the field limit of this arm is the
direct-path ratio. The four unequalised items of
Section~\ref{sec:arms} and the limitations of Section~\ref{sec:limits} bound
what these numbers support. The planned over-the-air capture at the surveyed site with
the mast and the vehicle surveyed will be used to anchor the geometry and the absolute level of the measurement.

\section*{Acknowledgment}
The channel traces were rendered with Sionna RT~\cite{sionna,sionna-rt}. The
live-downlink recording used for the delay-spread comparison in
Section~\ref{sec:limits} was published by Daniel Est\'evez under CC~BY~4.0 and
is used unmodified apart from truncation~\cite{estevez-lte-b20}; its author
is not associated with this work and did not review it. Building footprints are
\copyright{} OpenStreetMap contributors, used under the Open Database Licence;
the band-3 site records are from the Australian Communications and Media
Authority's Register of Radiocommunications Licences.

\bibliographystyle{IEEEtran}
\bibliography{references}
\end{document}